\documentclass[aps,prd,showpacs,twocolumn,superscriptaddress,floatfix]{revtex4}

\usepackage{amsmath,amssymb,graphicx,color}
\usepackage{hhline}

\newcommand{\be}{\begin{equation}}
\newcommand{\ee}{\end{equation}}
\newcommand{\bea}{\begin{eqnarray}}
\newcommand{\eea}{\end{eqnarray}}
\newcommand{\non}{\nonumber}

\begin{document}

\title{Particlelike solutions in modified gravity: the Higgs monopole.}

\author{Sandrine Schl\"ogel}
\email{sandrine.schlogel@unamur.be}
\affiliation{Namur Center for Complex Systems (naXys) \& D\'epartement de Math\'ematique, Universit\'e de Namur, \\
Rue de Bruxelles 61, B-5000 Namur, Belgium.}
\affiliation{Centre for Cosmology, Particle Physics and Phenomenology,
Institute of Mathematics and Physics, Louvain University, 
2 Chemin du Cyclotron, 1348 Louvain-la-Neuve (Belgium)}

\author{Massimiliano Rinaldi}
\email{massimiliano.rinaldi@unitn.it}
\affiliation{Namur Center for Complex Systems (naXys) \& D\'epartement de Math\'ematique, Universit\'e de Namur, \\
Rue de Bruxelles 61, B-5000 Namur, Belgium.}
\affiliation{Dipartimento di Fisica, Universit\`a di Trento \& TIFPA-INFN Trento,\\ Via Sommarive 14, 38123 Povo (TN), Italy.}

\author{Fran\c{c}ois Staelens}
\affiliation{Namur Center for Complex Systems (naXys) \& D\'epartement de Math\'ematique, Universit\'e de Namur, \\
Rue de Bruxelles 61, B-5000 Namur, Belgium.}

\author{Andr\'e F\"uzfa}
\email{andre.fuzfa@unamur.be}
\affiliation{Namur Center for Complex Systems (naXys) \& D\'epartement de Math\'ematique, Universit\'e de Namur, \\
Rue de Bruxelles 61, B-5000 Namur, Belgium.}
\affiliation{Centre for Cosmology, Particle Physics and Phenomenology,
Institute of Mathematics and Physics, Louvain University, 
2 Chemin du Cyclotron, 1348 Louvain-la-Neuve (Belgium)}

\date{\today}


\begin{abstract}
\noindent 
Higgs inflation has received a remarkable attention in the last few years due to its simplicity and predictive power. The key point of this model is the nonminimal coupling to gravity in unitary gauge. As such, this theory is in fact a scalar-tensor modification of gravity that needs to be studied also below the energy scales of inflation. Motivated by this goal, we study in great analytical and numerical detail the static and spherically symmetric solutions of the equations of motion in the presence of standard baryonic matter, called ``Higgs monopoles'' and presented in \cite{monopole}. These particlelike solutions may arise naturally in tensor-scalar gravity with mexican hat potential and are the only globally regular asymptotically flat solutions with finite classical energy. In the case when the parameters of the potential are taken to be the ones of the standard model, we find that the deviations from general relativity are extremely small, especially for bodies of astrophysical size and density. This allows to derive a simplified description of the monopole, for which the metric inside the spherical matter distribution can be approximated by the standard metric of general relativity. We study how the properties of these monopoles depend on the strength of the nonminimal coupling to gravity and on the baryonic mass and compactness. An important and original result is the existence of a  mechanism of resonant amplification of the Higgs field inside the monopole that comes into play for large nonminimal coupling. We show that this mechanism might degenerate into divergences of the Higgs field that reveal the existence of forbidden combinations of radius and baryonic energy density.
\end{abstract}
 


\maketitle


\section{Introduction}\label{intro}


\noindent Since the discovery of cosmic acceleration, the status of the theories of modified gravity has
drastically changed, from a pure theoretical possibility to a concrete explanation of effects that general
relativity (GR) alone cannot explain. The coincidence problem, together with naturalness issues, makes a
simple cosmological constant a very unlikely explanation for cosmic acceleration, opening the way to more
complex theories. However, modifying gravity is a dangerous business. Indeed, there are plenty of
phenomena that are very sensitive even to very small deviations from GR. Roughly
speaking, these can be divided into local tests of GR \cite{will}, mainly performed by
the help of Solar System observations (e.g.\ Cassini, \cite{aurel}) and cosmological tests, based on 
large surveys that probe the large-scale structure formations through the galaxy distributions at
different distances \cite{euclid}, and on the cosmic microwave background \cite{wmap}. To these,
one should add strong field, astrophysical effects, such as the ones related to pulsar timings.  To
be a robust theory, any model of modified gravity  must be compatible with the constraints
imposed by all of these tests.
            
Among the models of modified gravity, a prominent role is played by scalar-tensor theories, developed from
the pioneering work of Brans and Dicke \cite{brans}. On one hand, the scalar field represents the simplest
generalization of a cosmological constant to a dynamical dark energy component. On the other hand, some of
these models show interesting screening mechanisms that make them compatible with the severe constraints
imposed by Solar System observations (e.g. the chameleon, \cite{cham}). 

More recently, a great effort was paid to construct models where the same scalar field is responsible for
both inflation and late dark energy. In particular, the most general scalar-tensor theory with at most
second order equations of motion seems to achieve this result \cite{copeland}, although it brings along a
huge parameter space that still needs to be explored \cite{fabfour}.

Another promising line of research identifies the inflaton with the Higgs field nonminimally coupled to
gravity, according to the idea put forward in \cite{shaposh}. This model enjoys a striking simplicity that
explains very naturally the inflationary era without the need of non-standard model fields. Following concerns about a
possible loss of unitarity \cite{unitarity}, the model has been extended to include additional scalar
fields and unimodular gravity \cite{unimodular}.

In this paper we study in detail the so-called ``Higgs monopole'', namely the static, spherically symmetric, non-singular, 
and asymptotically flat particlelike solution 
that can be found when the Higgs field is nonminimally coupled to gravity and minimally coupled to ordinary baryonic matter. This solution was first described  in \cite{monopole}, where the numerical evidence of nontrivial spherically symmetric solutions with a large spontaneous scalarization
for certain values of the compactness and of the mass of the baryonic matter was presented. In the present paper, we would like to explore more properties of these new objects, with a comprehensive numerical and analytical study of  the
equations of motion. We begin in Sec.\ \ref{sec2} by presenting the model and the equations of motion. We
proceed in Sec.\ \ref{sec3}  by showing the mechanism that allows for the existence of the Higgs monopoles. In
Sec.\ \ref{sec4}, we analyze in detail some analytical properties of the equations of motion to prepare the
ground to the full numerical analysis. We also show an approximate solution to the Tolman-Oppenheimer-Volkoff (TOV) 
equation and we perform the post-netwonian analysis (PPN). We present the numerical study of the
Higgs monopoles in Sec.\ \ref{sec5} and we analyze the dependence on the mass, compactness and nonminimal coupling parameter. In particular, we show numerically that there exists a resonant amplification 
of the Higgs field magnitude inside the matter distribution for specific values of its radius. This effect is explained also analytically in Sec.\ \ref{sec6} and used to prove that there exists forbidden values for the radius of the compact object in the case of strong nonminimal coupling. We finally comment our results and draw some conclusions in Sec.\ \ref{conclusions}. In  appendix \ref{appendix1} we present an additional weak field analysis while in appendix \ref{appendix2} we report  the technical details of our numerical analysis.


\section{The Model}\label{sec2}


\noindent In this section we lay down the fundamental equations of the model. Let us begin with the Lagrangian of 
the Higgs field nonminimally coupled to gravity. In the Jordan frame, it
reads \footnote{We use the mostly plus signature and we set $c=1$.}
 
\bea
\mathcal{L}&=&\sqrt{g}\left[{F\left(H\right)\over {2\kappa}}{R}-\frac{1}{2}\left(\partial H\right)^2-V\left(H\right)\right]\\\non
&+&\mathcal{L}_{m}\left[g_{\mu\nu};\Psi_m\right],
\label{lagrangian}
\eea
where $H$ is the Higgs scalar field in the unitary gauge, $R$ is the
 Ricci scalar, $\Psi_m$ denotes  generic baryonic matter fields, and
  $\kappa=8\pi/m_{p}^{2}$, $m_p$  being the Planck mass. The potential $V$ has the usual mexican hat profile
\be\label{mexican}
V(H)={\lambda_{\rm sm}\over 4} (H^{2}-v^2)^2 ,
\ee
where $\lambda_{\rm sm}\sim 0.1$ \footnote{In this paper we use the standard model values for the parameters of the potential  $\lambda_{\rm sm}$ and $v$. In the future, we plan to study this theory as a generic scalar-tensor theory to see in which range these parameters are compatible with the current observations.} and $v=246$ GeV is the vacuum expectation value (vev) of the Higgs field. 
The nonminimal coupling function between the Ricci scalar and the scalar field is chosen to be the same as Higgs inflation, namely
  \be
F(H)=1+{\xi H^2\over m_p^2}, 
\ee
It is known that this model yields a successful inflation provided $\xi$ is large, of the order $10^{4}$ \cite{shaposh}. The form of this
 coupling function is further justified by invoking the (semiclassical) renormalization of the energy momentum tensor
  associated to the scalar field on a curved background, which needs terms like $H^{2}R$ in the Lagrangian
   \cite{renorm}. We will consider only positive values of $\xi$ to avoid the possibility that the effective reduced
    Planck mass (that can be identified with $(m_{p}^{2}+\xi H^{2})^{1/2}$) becomes imaginary.

A similar Lagrangian for compact objects was already considered in \cite{sudarsky}, where, however, the potential was
neglected. As we will see below, this is an important difference as the presence of the Higgs potential prevents the
solution to smoothly converge to GR. In other words, the solution $H=0$ does not yield the
Schwarzschild solution but, rather, a de Sitter black hole with a cosmological constant proportional to $v^{4}$.

It should also be kept in mind that the Higgs field is in general a complex doublet and, here, it is reduced to a
single real component by choosing the unitary gauge \cite{shaposh}. However, the other components, also known as
Goldstone bosons, can have physical effects, especially at high energy, when renormalizability imposes a different gauge choice (e.g.\ the so-called $R_{\xi}$-gauges, see for example \cite{peskin}). 
In cosmology, the effects of the Goldstone boson in a toy $U(1)$ model was  investigated by one of us in
\cite{maxhiggs}. In the context  of compact object, some results can be found in \cite{gleiser} although the
potential is not of the Higgs type.        

The equations of motion obtained from the Lagrangian
\eqref{lagrangian}  by variation with respect to the metric read
\be
 \left(1+{\xi\over m_p^2} H^2\right)G_{\mu\nu}=\kappa\left[T_{\mu\nu}^{(H)}+T_{\mu\nu}^{(\xi)}+T_{\mu\nu}^{(mat)}\right],
\label{eom_tensor}
\ee 
where $G_{\mu\nu}$ is the Einstein tensor, 
\bea\label{tmunuH}
T_{\mu\nu}^{(H)}&=&\partial_{\mu} H\partial_{\nu} H - g_{\mu\nu}\left[ \frac{1}{2} \left(\partial H\right)^2 + V\left(H\right)\right],
\eea
is the part of the stress-energy tensor associated to the Higgs field, and 
\be
T_{\mu\nu}^{(\xi)}=-{\xi\over 4\pi} \left[g_{\mu\nu} \nabla^\lambda\left(H \nabla_{\lambda} H \right)-\nabla_{\mu}\left(H \nabla_{\nu} H\right) \right],
\ee
is the stress-energy tensor induced by the nonminimal coupling $\xi$. Finally, 
\be
T_{\mu\nu}^{(mat)}={2 \over \sqrt{-g}} {\delta \mathcal{L}_{m} \over \delta g^{\mu\nu}},
\ee
is the stress-energy tensor of the baryonic matter fields that we assume to have the form of a perfect fluid, so that
\bea
T_{\mu\nu}^{(mat)}&=& \left(\rho+ p \right) u_\mu u_\nu +g_{\mu\nu} p,
\eea
where $u_\mu$ is the four-velocity, $\rho$ is the density and $p$ the pressure. Here, we adopt the splitting of the energy momentum tensor proposed in \cite{sudarsky} 
as each part will give distinct contributions, as we will see in section\ \ref{sec5}. 
The set of equations of motion is completed by the Klein-Gordon equation
\be
\Box H+\frac{\xi HR}{8\pi}=\frac{dV}{dH},
\label{KG}
\ee 
from which we can understand in a qualitative way the main characteristics of the solution, as we show in the next section.


\section{Effective dynamics}\label{sec3}


\noindent Our first goal is to assess whether spherically symmetric and asymptotically flat solutions to the equations of motion exist. 
The term in the Klein-Gordon equation \eqref{KG} that tells us if this is the case, is the one proportional to $\xi HR$. 
For a start, it is clear that $H=0$ is a trivial solution of the Klein-Gordon equation. If we consider a static and spherically symmetric spacetime, described by the metric 
\be
ds^{2}=-e^{2\nu\left(r\right)}dt^{2}+e^{2\lambda\left(r\right)} dr^{2}+r^{2} d\Omega^{2},
\label{schwar}
\ee
we see from \eqref{lagrangian} that, for $H=0$ and in the absence of matter, we obtain a de Sitter black hole solution because of the term proportional to $v^{4}$ in the potential \eqref{mexican}. Therefore, this solution is not asymptotically flat. 
In the absence of nonminimal coupling $(\xi=0)$ the solution $H=\pm v$ leads to the usual
Schwarzschild metric (with or without internal matter). On the other hand, with  a nonminimal
coupling and in the vacuum, there are no-hair theorems that force the solution to be the Schwarzschild one, i.e. again 
$H(r)=\pm v$ everywhere \cite{soti}. Therefore, the only interesting case is the one with nonminimal coupling and nonvanishing baryonic matter density, 
which, as we will show, has indeed finite energy and is asymptotically flat.  

To examine in detail the dynamics, we rewrite the Klein-Gordon equation \eqref{KG} as:
\be
\Box H=-{dV_{\rm eff}\over dH},
\ee
where 
\bea
V_{\rm eff}=-V+\frac{\xi H^2R}{16\pi}+\mathcal{C},
\label{Veff}
\eea
$\mathcal{C}$ being a constant of integration. Note that the form of the effective potential in a time-dependent inflationary background is quite different from the
form in a static and spherically symmetric one. In fact, if the metric has the flat Friedmann-Lema\^ itre-Robertson and Walker form
\be
ds^2=-dt^2+a(t)^2 \left(dr^2+r^2 d\Omega^2\right),
\ee
the scalar field rolls down (in time) into the potential well since the Klein-Gordon equation has the form
\be
{d^{2}H\over dt^{2}}+{3\over a}\frac{da}{dt}\frac{dH}{dt} = {dV_{\rm eff}\over dH}.
\ee
On the other hand, with the static and spherically symmetric  metric \eqref{schwar} the Klein-Gordon equation becomes
\bea
H''-H'\left(\lambda'-\nu'-\frac{2}{r}\right)&=&-{dV_{\rm eff}\over dH}\\
&=&\left(-\frac{\xi R}{8\pi }+\lambda_{\rm sm}(H^2-v^2)\right)H,\nonumber
\eea
where the prime denotes a derivative with respect to the radial coordinate $r$. For minimal coupling $\xi=0$, while $H=\pm v$ ($H=0$) 
corresponds to local minima (maximum) in the cosmological case, it corresponds to local maxima (minimum) in the spherical symmetric static configuration.  In addition, for nonminimal coupling, $H=0$ is
a stable equilibrium point while $H=v$ is an unstable one. In order to fully characterize the stability of these points in the nonminimal coupling case, it is necessary to compute $R$.

\begin{figure}
\begin{center}
\includegraphics[scale=0.35]{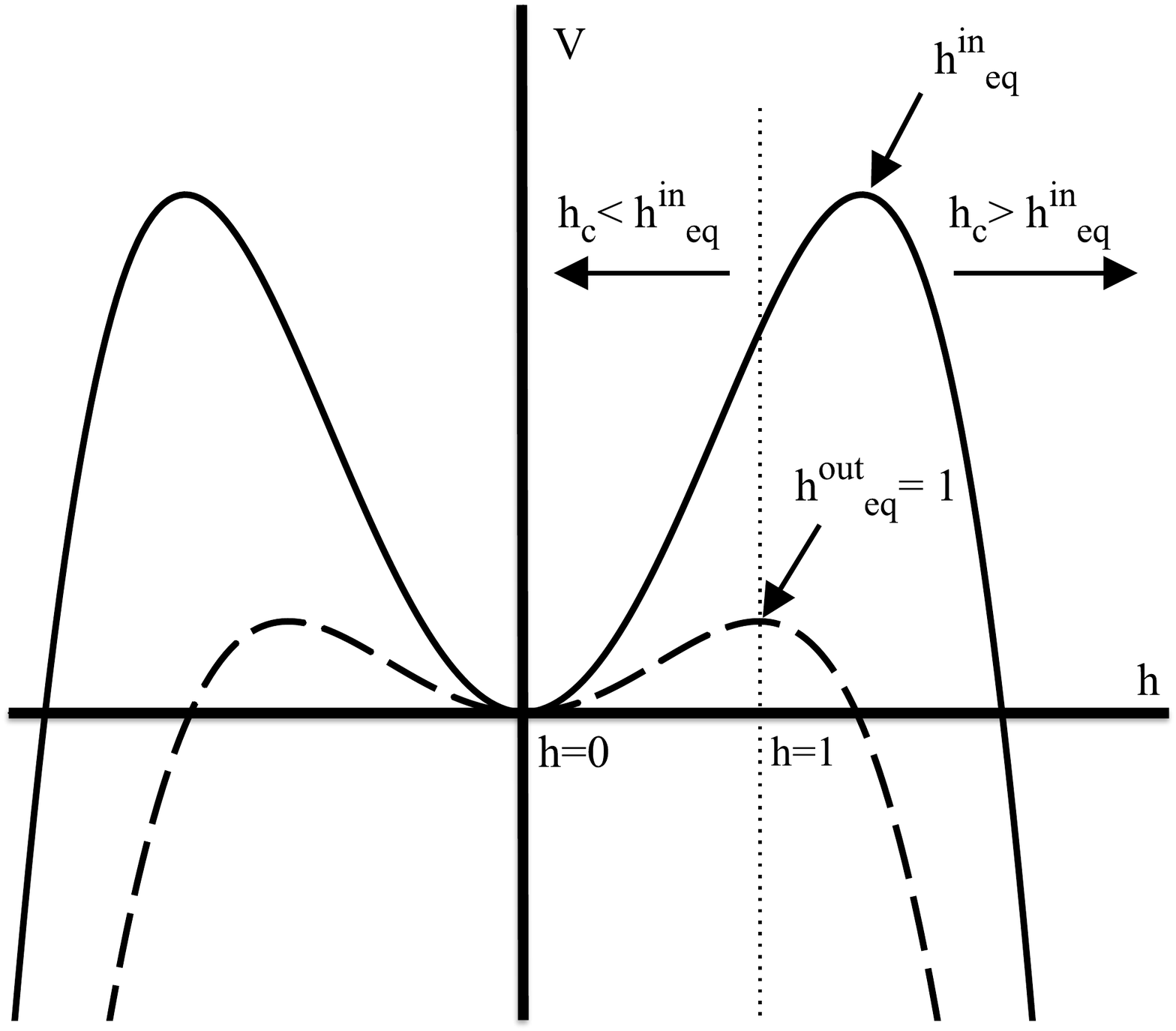}
\caption{
Qualitative plot of the potential inside (solid line) and outside (dashed line) the body. The effective potential
corresponds to the Higgs one outside the body while the local maxima (see $h^{\rm in}_{\rm eq}$ with $H=m_{Pl}vh$) are displaced from the vev inside the body.}
\centering
\label{plot_Veff2}
\end{center}
\end{figure}

For simplicity, from now on we will consider a top-hat distribution of baryonic matter, namely 
\bea\label{rho}
\rho(r)=\left\{\begin{array}{ll}\rho_{0} & 0<r<{\cal R}, \\0 & r>{\cal R},\end{array}\right.
\eea
where ${\cal R}$ is the radius of the spherical body. In this case, the effective potential shows a sharp transition 
between the interior and the exterior of the body. Indeed, if $\xi\neq 0$, the second term in Eq.\ \eqref{Veff} comes 
into play and we can show that the Ricci scalar satisfies the inequality
\be
R(r<\mathcal{R})\gg R(r>\mathcal{R}). 
\ee
The reason is that, inside the body, the Higgs field turns out to be almost constant, as we will see in Sec.\ \ref{sec5}. 
Therefore, all the derivatives in the trace of the energy-momentum tensor vanish and the only consistent 
contribution to $R$ comes from the potential, as one can easily check by calculating the trace of Eq.\  \eqref{eom_tensor}. If the 
Higgs field is not too much displaced from its vev inside the body, the greatest contribution to the curvature then 
comes from the baryonic matter, provided the density is sufficiently large.  Outside the body, the Higgs field drops 
very rapidly towards its vev and $R$ vanishes at large $r$ to match the Schwarzschild solution $R=0$ everywhere. For practical purposes, 
this means that we can approximate $R$, inside the body, as there was no Higgs field but just matter. To show this 
property a bit more rigorously, it is sufficient to calculate the trace of Eq.\ \eqref{eom_tensor} and recall that, at 
the center of the body, we must have $dH/dr=0$ according to standard  symmetry arguments. Therefore, near the center 
of the body, we can approximate the trace of Einstein equation as
\bea\label{apptrace}
\left(1+{\xi H^{2}\over m_{p}^{2}}\right)R\simeq -\kappa\left(2V+T^{(mat)}-{3\xi\over 4\pi}H\square H\right).
\eea
In addition, if we 
consider energies far below the Planck scale, $H\lll m_{p}$, all the terms like $\xi H^{2}/m_{p}^{2}$ can be safely 
neglected, even when $\xi\sim 10^{4}$. By also using the Klein-Gordon equation \eqref{KG}, we finally find that Eq.\ \eqref{apptrace} can be accurately approximated by
\bea\non
&&{1\over v^{2}}\left(R+\kappa T^{(mat)}\right)=-{\kappa\over v^{2}}\left(2V-{3\xi H\over 4\pi}{dV\over dH}\right)\\
&&=-{4\pi\lambda v^{2}\over m_{p}^{2}}\left({H^{2}\over v^{2}}-1\right)\left[{H^{2}\over v^{2}}\left(1-{3\xi\over 2\pi}\right)-1\right].
\label{approx_dyn}
\eea

Now, since $(v/m_{p})^{2}\sim 10^{-34}$, we need a very large ratio $H/v$ to yield a non-negligible right hand side, 
even for $\xi$ of the order of $10^{4}$. Therefore, unless we consider planckian energies for the Higgs field, the 
left hand side of the above equation is negligibly small, at least near the center of the body. This means that, 
inside the body, the Einstein equation is undistinguishable from the standard GR equation $R=-\kappa T^{(mat)}$. In the appendix B, we will show numerically that this approximation is accurate and can be 
used to 
investigate the monopoles for a very large range of parameters.

Let us now study the equilibrium points of $V_{\rm eff}$. Outside the body, where $T^{(mat)}=0$, we can approximate 
$R\simeq 0$. Therefore, $dV_{\rm eff}/dH$ vanishes at 
\be
{H^{\rm out}_{\rm eq}\over v}=0,\,\pm 1.
\ee 
Inside the body, where $R\simeq -\kappa T^{(mat)}$, we find instead
\be
{H^{\rm in}_{\rm eq} \over v}=0,\,\pm \sqrt{1+{R \xi \over 8\pi\lambda_{\rm sm} v^2}},
\ee
and the crucial role of a nonvanishing $\xi$ becomes evident. 

The solutions that we are looking for, must interpolate between 
the value of the Higgs field at the center of the body $H_c$  and at the spatial infinity $H=\pm v$.  
Furthermore, since  $H'$ must vanish at the origin,  $H$ rolls down 
into the effective potential from rest. Suppose that $\vert H_c\vert$ is greater than the nonzero value of $\vert 
H^{\rm in}_{\rm eq}\vert$. Then, the Higgs field will roll outwards increasing boundlessly its value without any possibility 
of reaching an equilibrium outside the body. On the contrary, if $\vert H_c \vert $   is smaller than the nonzero 
value of $\vert H^{\rm in}_{\rm eq}\vert$, the Higgs field rolls down inward, towards the equilibrium at $H_\infty=0$, see 
Fig.\ \ref{plot_Veff2}.  For a given energy density and radius of the body, there is only one initial value 
$H_{c}$ such that $H(r)$ smoothly rolls towards $H_{\infty}=v$ and the geometry is asymptotically flat. All the other 
trajectories lead to either an asymptotically
de Sitter solution or to a divergent Higgs field at infinity. These particular solutions, with finite energy and 
asymptotically flat geometry,
are dubbed \emph{Higgs monopoles} and will be numerically obtained with a specifically designed shooting method in 
the following sections.


\section{Analytic properties}\label{sec4}


\noindent Before exploring the numerical solutions to the equations of motion, it is worth investigating their analytical properties  in order to obtain information able to target  more efficiently the numerical analysis. We find more convenient, for this section, to 
write the Lagrangian \eqref{lagrangian} in the standard Brans-Dicke form 
\bea\label{BDaction}
{\cal L}_{BD}={\sqrt{g}\over 2\kappa}\left[\phi R-{\omega\over \phi}(\partial \phi)^{2}-\bar V(\phi)\right]+{\cal L}_{m}.
\eea
where
\bea
\phi=1+{\xi H^{2}\over m_{p}^{2}},\quad \omega(\phi)={2\pi\phi\over \xi(\phi-1)},
\eea
and
\bea
\bar V(\phi)={\kappa\lambda_{\rm sm}\over 2}\left[{8\pi\over \xi\kappa}(\phi-1)-v^{2}\right]^{2}.
\eea
The Einstein equations now read
\bea\label{ee}
R_{\mu\nu}-{1\over 2}g_{\mu\nu}R&=&{1\over \phi}\nabla_{\mu}\nabla_{\nu}\phi+{\omega\over\phi^{2}}\nabla_{\mu}\phi\nabla_{\nu}\phi \\\non
&-&{1\over\phi}\left[\square\phi+{\omega\over 2\phi}(\partial\phi)^{2}+{\bar V\over 2}\right]g_{\mu\nu}+{\kappa\over\phi}T_{\mu\nu},
\eea
where $T_{\,\,\mu}^{\nu}={\rm diag}(-\rho,p,p,p)$ is the energy momentum tensor of the fluid. Using the trace of this 
equation, we can write the Klein-Gordon equation as
\bea
(2\omega+3)\square\phi+{d\omega\over d\phi}(\partial\phi)^{2}-\phi{d\bar V\over d\phi}+2\bar V= \kappa T.
\eea
With the metric \eqref{schwar}, the $tt-$ and $rr-$components of the Einstein equations are, respectively
\begin{widetext}
\bea
\lambda'\left({2\over r}+{\phi'\over\phi}\right)-{\kappa\rho\over \phi }e^{2\lambda}+{1\over r^{2}}\left(e^{2\lambda}-1\right)-{\phi''\over\phi}-{2\phi'\over r\phi}-{\bar V e^{2\lambda}\over 2\phi}
-{\omega\over 2}\left( \phi'\over \phi\right)^{2}&=&0,\\
\nu'\left({2\over r}+{\phi'\over\phi}\right)-{\kappa p\over\phi }\,e^{2\lambda}-{1\over r^{2}}\left(e^{2\lambda}-1\right)+{2\phi'\over r\phi}+{\bar V e^{2\lambda}\over 2\phi}
-{\omega\over 2}\left( \phi'\over \phi\right)^{2}&=&0,
\eea
while the angular component is
\bea
\nu''+(\nu')^{2}+\nu'\left({1\over r}+{\phi'\over \phi}\right)-\lambda'\left(\nu'+{1\over r}+{\phi'\over\phi}\right)+{\phi''\over\phi}+{\phi'\over r\phi}+{\omega\over 2}\left( \phi'\over \phi\right)^{2}+{e^{2\lambda}\over\phi}\left({\bar V\over 2}-\kappa p\right)=0.
\eea
Finally, the Klein-Gordon equation becomes
\bea
(2\omega+3)\left(\phi''+\phi'\nu'-\phi'\lambda'+{2\over r}\phi'\right)+(\phi')^{2}{d\omega\over d\phi}+e^{2\lambda}\left(2\bar V-\phi{d\bar V\over d\phi}\right)+\kappa e^{2\lambda}(\rho-3p)=0.
\eea
\end{widetext}
To these we must add  the relation $p'+(p+\rho)\nu'=0$ obtained by the usual Bianchi identities. The total energy 
momentum tensor is identified with the right hand side of Eq.\ \eqref{ee}. Therefore, the total energy density, given 
by $\rho_{\rm tot}=-T^{0}_{\rm tot\,\,0}$, reads
\bea
\rho_{\rm tot}=e^{-2\lambda}\!\!\left[{\phi''\over \phi}-{\phi'\lambda'\over \phi}+{2\phi'\over r\phi }+{\omega 
\phi'^{2}\over 2\phi^{2}}\right]+{\bar V\over 2\phi}+{\kappa\rho\over\phi}.\label{rhotot}
\eea
As mentioned in the previous section, if $\phi$ (and hence $H$)  varies very slowly with $r$, the energy density is dominated by 
the baryonic matter. This is certainly true near the center, as there we must have $\phi'(r=0)=0$, as required by 
symmetry arguments.  As a consequence, all the derivatives are negligible and we are left with
\bea
\rho_{\rm tot}\simeq {\bar V_{c}\over 2\phi_{c}}+{\kappa\rho\over\phi_{c}},
\eea
where a subscript ``$c$'' indicates the value of a quantity at the center of the body. If $\bar V_{c}$ is not too 
large, that is the Higgs field is not displaced too much from its vev, then the energy density can be taken as the 
one of GR, as we have  already  shown in section\ \ref{sec3}.

Now, consider the Klein-Gordon equation and suppose that there exists a point $\bar r$ such that $\phi'(\bar r)=0$. 
Suppose also that the energy density is constant inside the body, $\rho=E$. It follows that, at that point
\bea
\bar\phi''= {e^{2\bar\lambda}\over (2\bar \omega+3) }\left[ {64\pi^{2}\lambda_{\rm sm} \phi_{v}(\bar\phi-\phi_{v})\over\kappa \xi^{2}}-\kappa(E-3\bar p)\right],
\eea
where the bar denotes quantities calculated at $\bar r$ and $\phi_{v}=1+\xi\kappa v^{2}/(8\pi)$ is the vev  of 
$\phi$. Outside the body, where $p=E=0$ everywhere and $e^{2\lambda}$ is positive definite, we have two cases:
\begin{itemize}
\item $\bar\phi''>0$, i.e. a local minimum, which implies that $\bar\phi>\phi_{v}$,
\item $\bar\phi''<0$, i.e. a local maximum, which implies that $\bar\phi<\phi_{v}$.
\end{itemize}
This shows that if there is a local maximum or a local minimum for $\phi$ at a point outside the body, then the field 
cannot converge to its vev $\phi_{v}$ at infinity. This is possible only if $\phi$ is a monotone and decreasing 
function of $r$ (or if $\phi=\phi_{v}$ everywhere). As we will see further, 
this property allows to approximate the Higgs field outside the body with a Yukawa function and an associated scalar charge. This is no longer true inside the body, as $E-3\bar p>0$ and the displacement of $\phi$ from its vev can be 
compensated by contributions from the energy density and the pressure. Thus, we can have both minima and maxima of the field inside the body. In other words, $\phi$ can oscillate only inside the body. The monopoles that will be presented in section \ref{sec5} will illustrate this analytical property.

Let us look at the equations of motion in the absence of matter, i.e. with $\rho=p=0$. 
From the Klein-Gordon equation, we see that, in this case, asymptotic flatness (namely $\lambda'= \nu'= \phi'\simeq0$ for large $r$) is consistent with $\phi(r\rightarrow\infty)=\phi_{v}\approx 1$ (since $\kappa\xi v^2\ll 1$)  only if  
$V(\phi_{\infty})-{1\over 2}V'(\phi_{\infty})=0$. As discussed in \cite{soti}, this 
implies that the only asymptotically flat solution to the equations of motion is the one that coincides with GR, namely the Schwarzschild metric with constant scalar field.

\subsection{Classical energy}

\noindent In scalar-tensor theories, it is customary to calculate the binding energy of the system and compare it to 
the GR value in order to see if a solution is energetically favoured. The binding energy is defined by the difference 
between the baryonic energy (the energy of the baryons if they were dispersed) and the ADM energy $E_{\rm bin}=E_{\rm bar}-E_{\rm ADM}$. The baryonic energy is defined by
\bea\non
E_{\rm bar}=\int_{V}d^{3}x\sqrt{{}^{(3)}g}\,n(r)m_{b}={4\pi\over \phi }\int_{0}^{\cal R}r^{2}\rho(r)e^{\lambda(r)}dr,\\
\eea
where $n(r)$ is the density number, $m_{b}$ is the average mass of a baryon, $\rho(r)$
is the density profile and $\sqrt{{}^{(3)}g}$ is the 
proper volume measure. The ADM energy is defined as
\bea\non
E_{\rm ADM}=-\!\!\int d^{3}x\sqrt{{}^{(3)}g}\Bigg|_{r=\infty}\!\!\!\!g_{tt}T^{tt}_{\rm tot}=-4\pi\!\int_{0}^{\infty}\!\!r^{2}\rho_{\rm tot}dr,\\
\eea
where $\rho_{\rm tot}$ is the total energy density, including the scalar field contributions. In our case, it is given by Eq.\ \eqref{rhotot}. 
In general, when the potential is such that the scalar field vanishes at its minimum, there are always two types of 
solutions. The first has a vanishing scalar field everywhere and coincides with standard GR solutions. In 
the absence of matter and angular momentum, this solution is the Schwarzschild metric. The second 
solution has a varying scalar field and it approaches the Schwarzschild solution only at spatial infinity. The 
important point is that the two families of solution are smoothly connected and this allows to compare the binding 
energy of the two configurations and to determine the stable one, or at least the one that minimizes the energy 
\footnote{The stability under small perturbations of the metric is a different and much more complicated issue that 
will not be considered in this work.}. In our case, however, this comparison  is meaningless since the monopole solution
cannot smoothly reduce to the Schwarzschild one because of the nonminimal coupling. In fact, the monopole is the 
unique solution (for a given energy density and radius) with finite energy. All other solutions have either a 
diverging or vanishing scalar field $H$ at spatial infinity, a result we have already established in section \ref{sec3}. In the first case, the potential term diverges so 
$E_{\rm ADM}$ is infinite. In the second case, if $H\rightarrow 0$ then $\phi\rightarrow (2\kappa)^{-1}$ so the term 
$r^{2}V/\phi$ diverges, yielding again an infinite ADM energy \footnote{For a correct calculation of the mass associated to a de Sitter black hole see Ref.\ \cite{GH}.}.

\subsection{TOV equation}

\noindent We now find an approximate formula for the pressure as a function of the energy density and the value of 
the scalar field at the center $\phi_{c}$, in analogy with the usual TOV equation. To do so, it 
is sufficient to expand and solve the equations of motion around $r=0$. It should be kept in mind that, for the 
monopoles, the value of $\phi_{c}$ (or, equivalently, of $\bar V(r=0)$) is not arbitrary. As we explained in the 
previous sections, the value of $\phi_{c}$ for a given mass and body radius is determined by the condition that 
$\phi=\phi_{v}$ at spatial infinity and, therefore,  cannot be fixed by a local expansion. This is the reason why the 
best we can do, analytically, is to find the central pressure $p_{c}=p(r=0)$ as a function of $\phi_{c}$. As before,
we assume that $\rho=E=$ const. At the center, owing to spherical symmetry, the scalar field can be approximated 
as $\phi\simeq \phi_{c}+\phi_{2}r^{2}$ so that we can solve the $tt$-component of the Einstein equation and find 
\bea
e^{2\lambda(r)}=  \left[1-{2m(r)\over r}\right]^{-1},
\eea
where
\bea\label{mofr}
m(r)\simeq {\left(2\kappa E+\bar V_{c}\right)r^{3}\over 12\phi_{c}}.
\eea
This result can be inserted in the $rr$-
component of the Einstein equation, which, together with the equation of state $p'+(p+\rho)\nu'=0$, gives the modified TOV equation
\bea
{dp(r)\over dr}\simeq-{(p(r)+E)(\kappa E+3 \kappa  p(r)- \bar V_{c})r\over 6\phi_{c}-(2\kappa  E+\bar V_{c})r^{2}},
\eea
that can be solved by separation of variables with the boundary condition that $p=0$ at $r={\cal R}$. The result reads
\bea\non
p(r)={E(\kappa E-\bar V_{c})\left(\sqrt{1-{2m/ r}}-\sqrt{1-{2m{\cal R}^{2}/ r^{3}}}\right)\over 3\kappa 
E\sqrt{1-{2m{\cal R}^{2}/r^{3}}}- (\kappa E-\bar V_{c})\sqrt{1-{2m/r}}},\\
\eea
where $m$ is the function \eqref{mofr}. This equation reproduces the relativistic expression in the limit  
$\bar V_{c}\rightarrow0$. At the center of the body we have
\bea
p_{c}={E(\kappa E-\bar V_{c})\left(1-\sqrt{1-{2m({\cal R})/ {\cal R}}}\right)\over {3 \kappa E}\sqrt{1-{2m({\cal 
R})/{\cal R}}}- \kappa E+\bar V_{c}},
\eea
where $m({\cal R})$ is the mass function \eqref{mofr} calculated at $r={\cal R}$. As for the ordinary relativistic 
stars, there is a maximum value of the energy density, at which the pressure diverges, given by 
\bea
E_{max}={{12\phi_{c}-\bar V_{c}{\cal R}^{2}}\over 2\kappa {\cal R}^{2}}.
\eea
However, in contrast with the GR case, there exists also a critical value of the energy density, below which the 
pressure becomes negative, that is
\bea
E_{min}={\bar V_{c}\over \kappa}&=&{\lambda_{\rm sm}\over 2}\left[ {8\pi\over \xi \kappa}(\phi_{c}-1)-v^{2} \right]^{2}, \\\nonumber
&=&{\lambda_{\rm sm}\over 2}\left( H_{c}^2-v^{2} \right)^{2}.
\eea
The interpretation is that the Higgs field contributes with a negative pressure at the center of the body, at least 
in the linearized regime considered in this section. When this approximation is no longer valid, we need to resort to 
numerical tools to calculate the central pressure and verify in which part of the parameter space it is negative and 
an eventual threat to the stability of the spherical body.


\subsection{PPN analysis}\label{PPN}

\noindent The Brans-Dicke formalism used in this section leads straightforwardly to the PPN analysis, which tells us 
the amount of deviations from GR outside a body of the size of the Sun. According to the PPN prescriptions, we 
assume that far outside the Sun, the Higgs field is close to its vacuum value so that $V\simeq 0$ and the Newton 
constant coincides with its bare value. The PPN parameters follow immediately and read
\bea
\gamma={\omega+1\over \omega+2},\quad \beta-1={1\over(2\omega+3)^{2}(2\omega +4)}{d\omega\over d \phi}.
\eea
When $\phi\rightarrow \phi_{v}$ it is not difficult to see that $\beta-1=\gamma-1=0$ with a precision far larger that 
actual observational constraints, as $\omega(\phi=\phi_{v})\simeq 1.5\times 10^{26}$ and $(d\omega/d\phi)
(\phi=\phi_{v})\simeq -2.2\times 10^{-20}$. Therefore, the theory is undistinguishable from GR at the 
level of Solar System experiments, even for large values of $\xi$. This result is consistent with the fact that, for a mass of the size and compactness of the Sun, 
the spontaneous scalarization is extremely small, as we will show in the next section. An alternative weak-field analysis is discussed in appendix \ref{appendix1}.


\section{Numerical results}\label{sec5}


\noindent After discussing the dynamics of the model and some generic analytical results, we now study numerically  the 
properties of the solutions. We report the reader to Appendix \ref{appendix2} for the complete set of equations of
motion, the system of units and the numerical methods that we used. In the previous sections, we have shown that the metric components inside the compact object are almost the same as in GR when we choose the standard model values for the parameters of the potential.
Therefore, we follow a simplified procedure, which consists in using the GR solution (with the top-hat matter distribution \eqref{rho}) for the metric components and focus solely on the non-trivial dynamics of the Higgs field. We provide for a proof of this approximation in Appendix B through the comparison between this approach and the numerical integration of the unaltered system of equations of motion.  With these assumptions, the problem  essentially reduces to solving the Klein-Gordon equation
\bea\non
h_{uu}+h_u \left(\nu_u-\lambda_u+\frac{2}{u}\right)\\
=e^{2\lambda}\left(-\frac{R \xi h}{8\pi}+\frac{r_s^2}{m_{pl}^2
\tilde v^2}\frac{dV}{dh}\right),
\label{kg_simp}
\eea
where $h=H/(m_{pl} \tilde{v})$, $\tilde{v}=v/m_{pl}$ being the dimensionless vev, and a subscript $u$ denotes a derivative with respect to $u=r/r_s$. Here,
$r_s=8\pi\rho_0/(3m_{Pl}^2) \mathcal{R}^3$ is nothing but the standard Schwarzschild radius \footnote{The physical Schwarzschild radius should take into account also the contribution of the Higgs field. Here, we define it instead as a scale of the theory, uniquely determined by the baryonic mass of the monopole as in GR.}. The metric fields and the scalar curvature are approximated by the standard Schwarzschild solution and read, respectively
\bea
\label{nu}\non
e^{2\nu}(u)&=&\left\{\begin{array}{lr}\frac{3}{2}\sqrt{1-s}-\frac{1}{2}\sqrt{1-s^3u^2}, &  \;0<u<s^{-1}, \\ \\ 1-u^{-1}, &
u\ge s^{-1},\end{array}\right.\\\\
\label{lam}
e^{-2\lambda}(u)&=&\left\{\begin{array}{lr} 1-s^3u^2, &  \qquad 0<u<s^{-1}, \\\\ 1-u^{-1}, & \qquad u\ge s^{-1},\end{array}\right.\\\non\\
\label{scalR}\non
R(u)&=&\left\{\begin{array}{lr} -\frac{6s^3}{r_s^2}\left(\frac{2\sqrt{1-s^3u^2}-3\sqrt{1-s}}{3\sqrt{1-s}-\sqrt{1-
s^3u^2}}\right), &  \;0<u<s^{-1}, \\\\ 0, & u\ge s^{-1},\end{array}\right.\\
\eea
where $s=r_s/\mathcal{R}$ is the compactness. Regularity at the origin requires that $h_u|_{u=0}=0$ leaving $h_c=h(u=0)$ as the only initial condition for Eq. (\ref{kg_simp}).

\begin{figure}[ht]
\begin{center}
\includegraphics[scale=0.30,trim=320 0 0 0,clip=true]{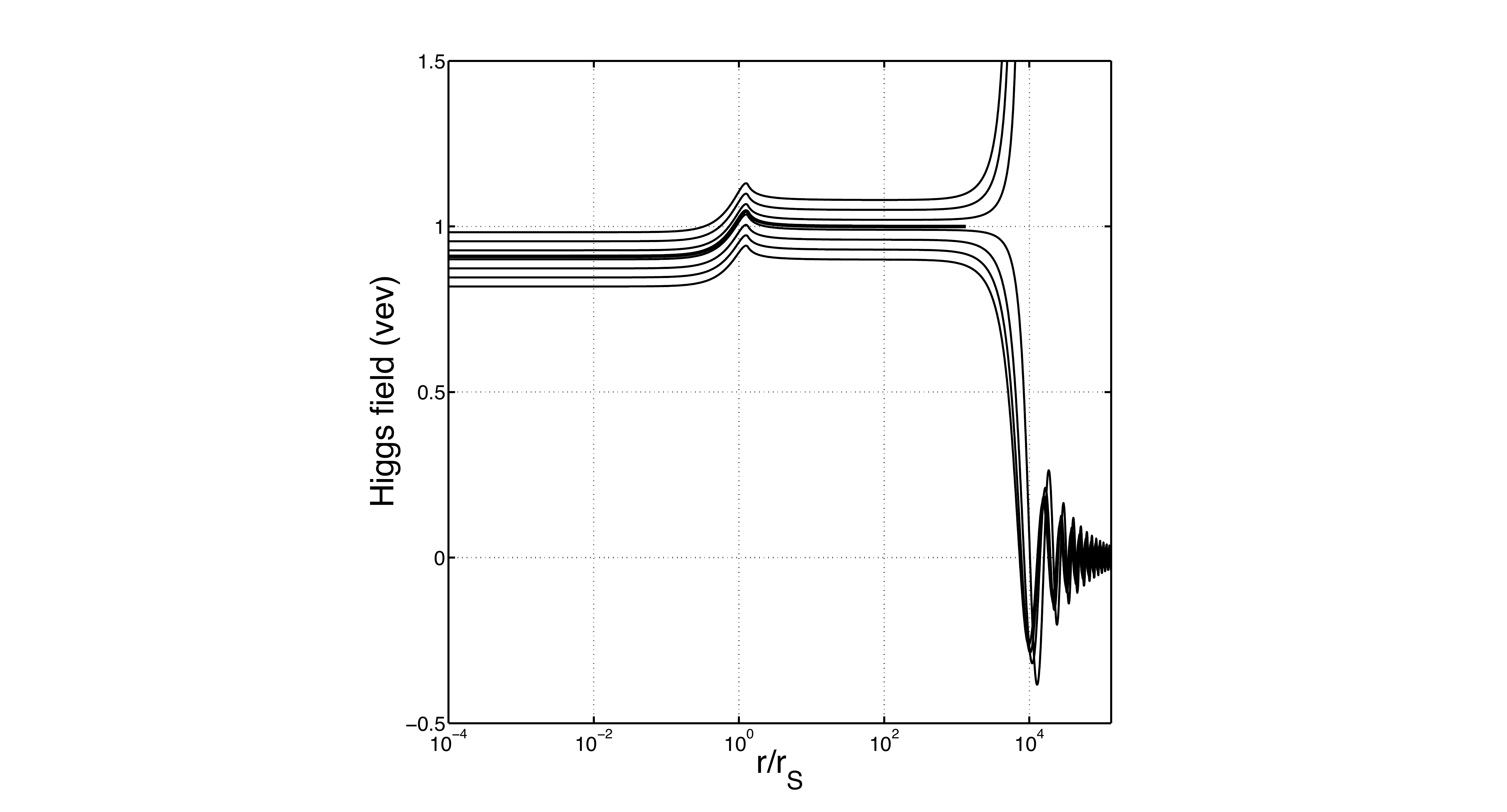}
\caption{Numerical solutions of Eq.\ \eqref{kg_simp} with varying initial conditions $h_c=h(r/r_s=0)$ for $\xi=10$, $m=10^6\; \rm kg$, and $s=0.75$. The thicker line represents the unique solution that converges to $h=1$ at large $r/r_{s}$.}
\centering
\label{different_hcs}
\end{center}
\end{figure}

In Fig.\ \ref{different_hcs} we plot the numerical solutions of Eq.\ (\ref{kg_simp}) for different values of the initial condition $h_c=h(u=0)$. We see that, for fixed mass and compactness, there exists only one value for the initial condition $h_c=h_0$ that yields a solution that tends to $h=1$ at spatial infinity (marked by a thicker line). This  solution corresponds to the non-trivial, asymptotically flat, and spherically symmetric distribution of the Higgs field, which was named ``Higgs monopole'' in \cite{monopole}. For 
slightly different initial conditions $h_c\ne h_{0}$, the field either diverges (if $h_c>h_0$) or tends to zero 
after some damped oscillations (if $h_c<h_0$). This result confirms the analytic treatment of Sec.\ \ref{sec3}.

\begin{figure}[ht]
\begin{center}
\includegraphics[scale=0.29, trim=340 0 350 0,clip=true] {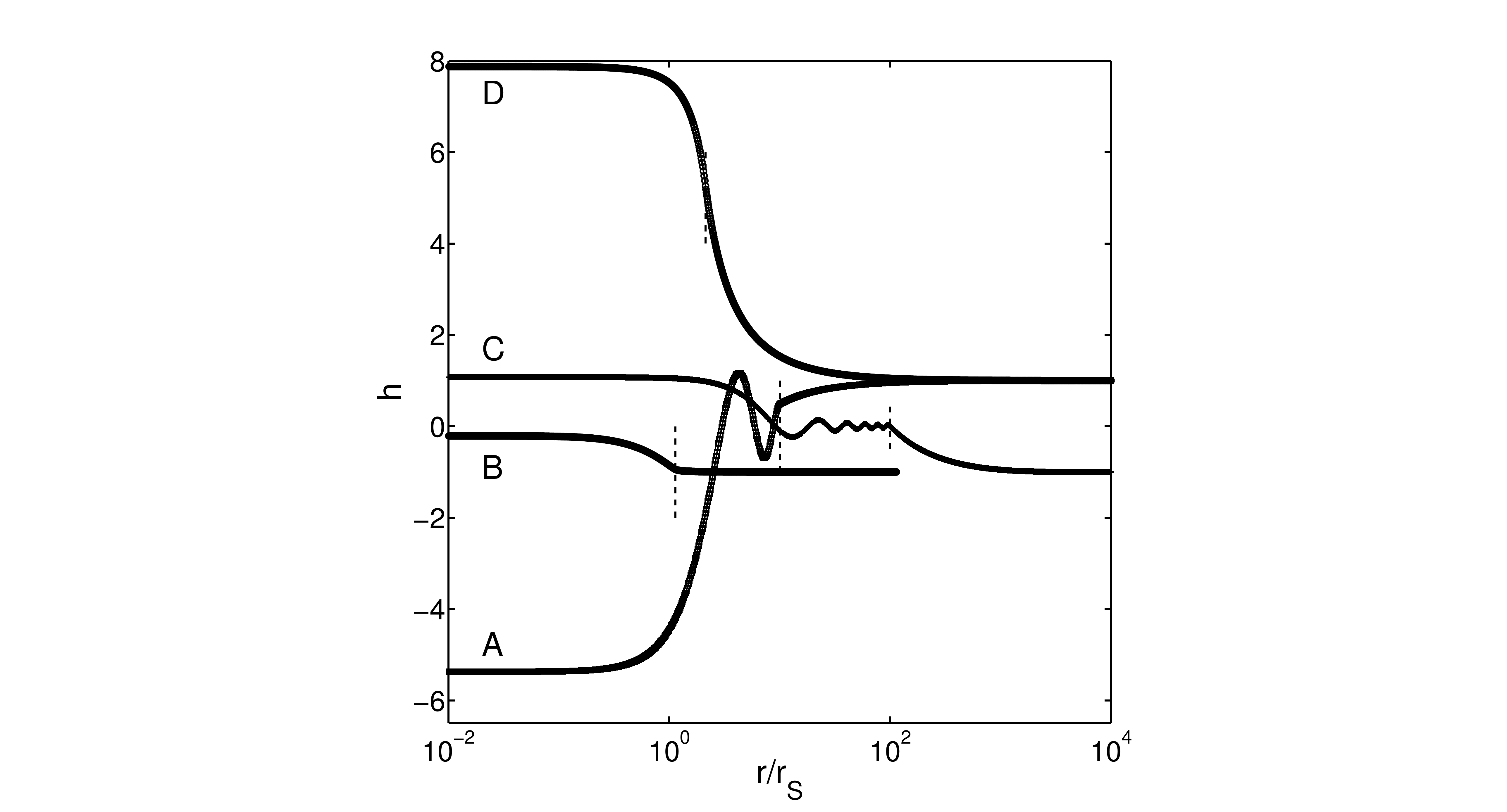}
\caption{Plots of the Higgs field with the parameters listed in Table I. The vertical dashed lines mark the radius of the body for each monopole.}
\centering
\label{plot_monop_family}
\end{center}
\end{figure}

For each choice of mass, compactness, and coupling strength $\xi$, there exists only one solution of the kind depicted in Fig.\ \ref{different_hcs}. Its form varies a lot in function of the parameters, as we show in Fig.\ \ref{plot_monop_family} where we plotted several solutions, corresponding to the parametrization listed in Table \eqref{table}. We notice that the value of the Higgs field at the center of the monopole 
can be lower than the vev for typically large compactness $s$. For small or moderate compactness, the 
central value of the Higgs field is generically larger that the vev. 

\begin{table}[htp]
\begin{tabular}{|c|c|c|c|c|c|}
\hline
 & $h_c$ & $\xi$ & m &  s \\
 \hline
 F & 0.91 & 10 & $10^6$ kg & 0.75 \\
 \hhline{|=|=|=|=|=|}
A & - 5.37 & $10^4$ & $10^3$ kg & $0.1$ \\
\hline
B &- 0.21 & $10$ & $10^6$ kg & $0.88$ \\
\hline
C & 1.077 & $10^6$ & $10^6$ kg & $0.01$ \\
\hline
D & 7.88 & $60$ & $10^4$ kg & $0.47$ \\
\hline
\end{tabular}
\caption{Properties of the Higgs monopoles plotted in Fig.\ \ref{different_hcs} (curve F) and Fig.\ \ref{plot_monop_family} (curves A,B,C,D).}\label{table}
\end{table}

\begin{figure}
\begin{center}
\includegraphics[scale=0.37,trim=280 0 310 0,clip=true]{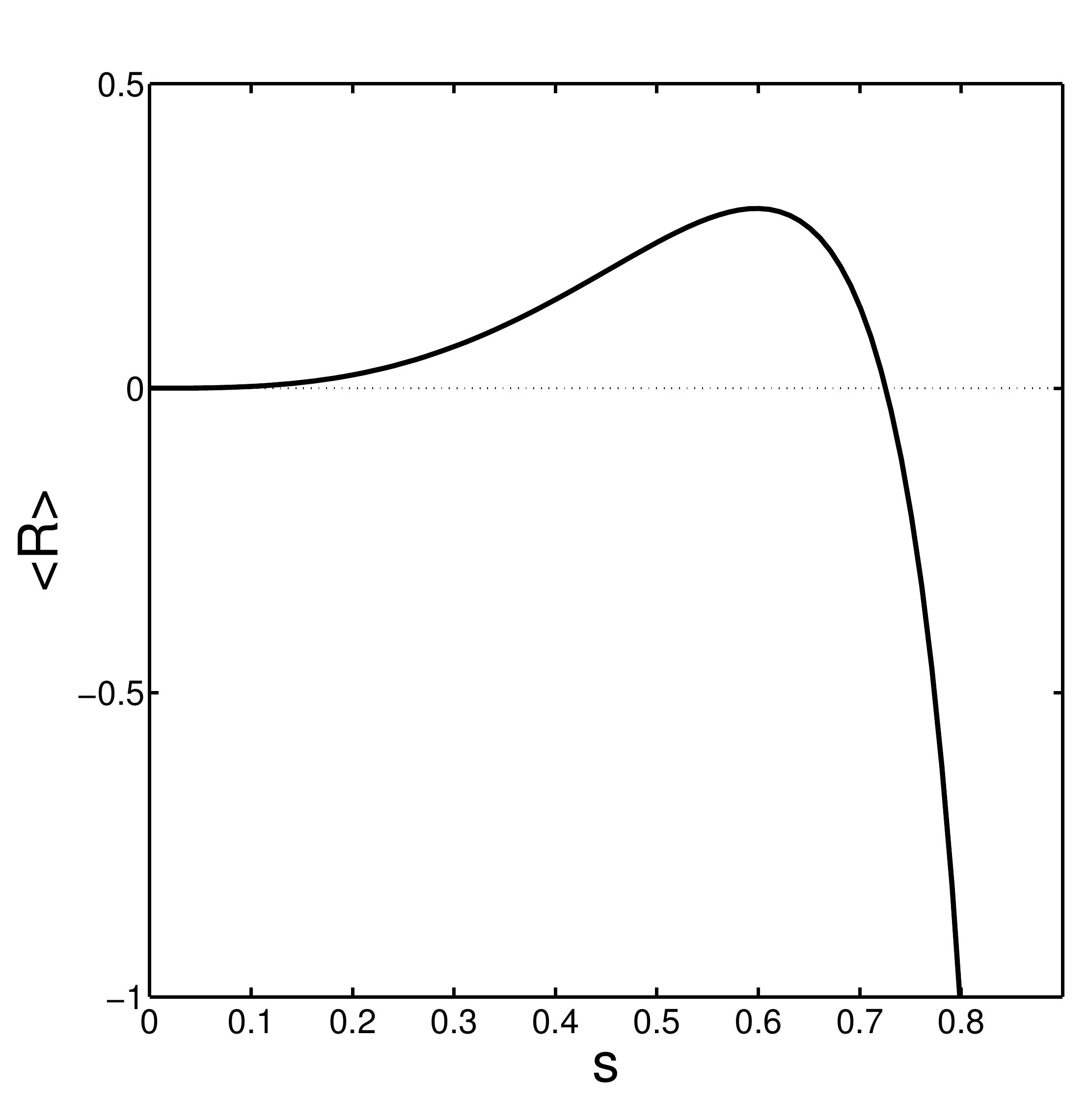}
\caption{Plot of $\langle R\rangle$ as a function of the compactness in the interval $[0, \cal R]$.}
\label{Rmean}
\end{center}
\end{figure}

Such behavior can be easily understood by considering the upper bound for $|h_c|$ introduced in
section \ref{sec3}. This upper limit corresponds
to the maxima of the effective potential inside the matter distribution. If we work in GeV units, we can express it as
\bea
h^{\rm in}_{\rm eq}=0,\pm \sqrt{1+{R \xi \over 8\pi\lambda_{\rm sm} \tilde{v}^2}}=0,\pm \sqrt{1+{R \xi \over 8\pi m_H^2}},
\label{h_eq_in}
\eea
where $m_H$ is the mass of the Higgs field. Since $R$ depends 
on the radial coordinate (see Eq.\ \eqref{scalR}) so does the effective potential. In order to show that we may have $|h_c|<1$, 
we approximate $R$ in Eq.\ \eqref{h_eq_in} by its spatial average 
\bea
\langle R\rangle=\frac{\int R(u)\sqrt{g}d^3x}{\int\sqrt{g}d^3x}=\frac{\int_0^{1/s} R(u) u e^{\lambda}du}{\int_0^{1/s} u e^{\lambda}du}.
\label{eqRmean}
\eea
In Fig.\ \ref{Rmean} we plot the value of  $\langle R\rangle$ in function of the compactness. We see that, for  $s\gtrsim 0.72$, $\langle R\rangle$ becomes negative so that $h^{\rm in}_{\rm eq}<1$, which implies  $|h_c|<1$. This happens, for instance, for the monopole represented by the curve B in Fig.\ \ref{plot_monop_family}.
In this plot we also notice that, for large $\xi$, oscillations 
are present only inside the compact body (see monopoles A and C), confirming the analytical results found in Sec.\ \ref{sec4}. 

\begin{figure}
\begin{center}
\includegraphics[scale=0.34,trim=370 10 90 10,clip=true]{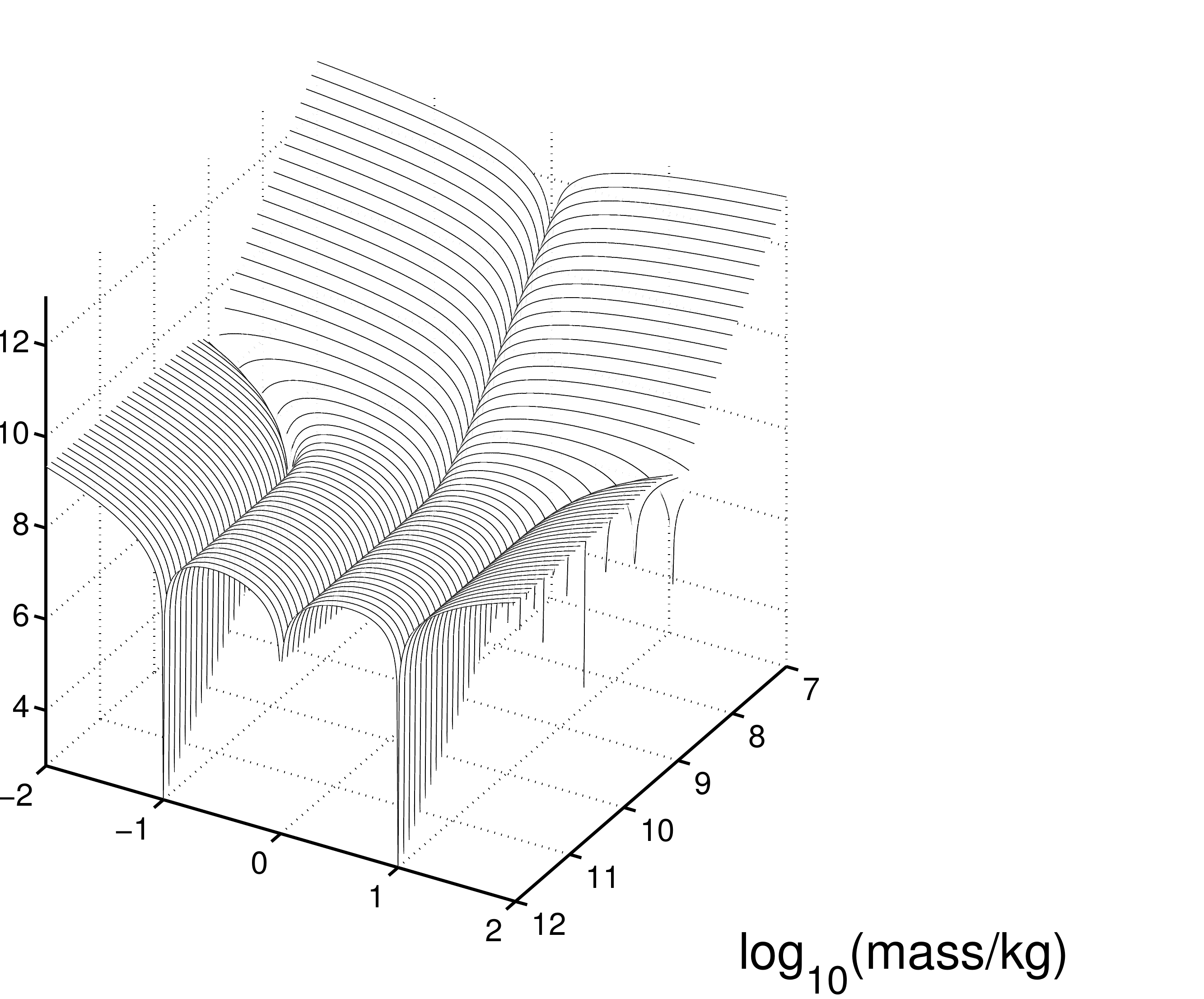}
\caption{
Derivative of the effective potential  $V_{\rm eff}$ of  Eq.\ \eqref{Veff} as a function of the mass of the monopole 
for fixed nonminimal coupling and compactness ($\xi=60$ and $s=0.5$). }
\label{dVeffdh}
\end{center}
\end{figure}

Finally, we point out that the central value of the Higgs field can be significantly larger than the vev (see e.g.\ the monopole D). We will see below that there is a novel amplification mechanism that explains these large values.
\begin{figure}
\begin{center}
\includegraphics[scale=0.37,trim=280 0 300 0,clip=true]{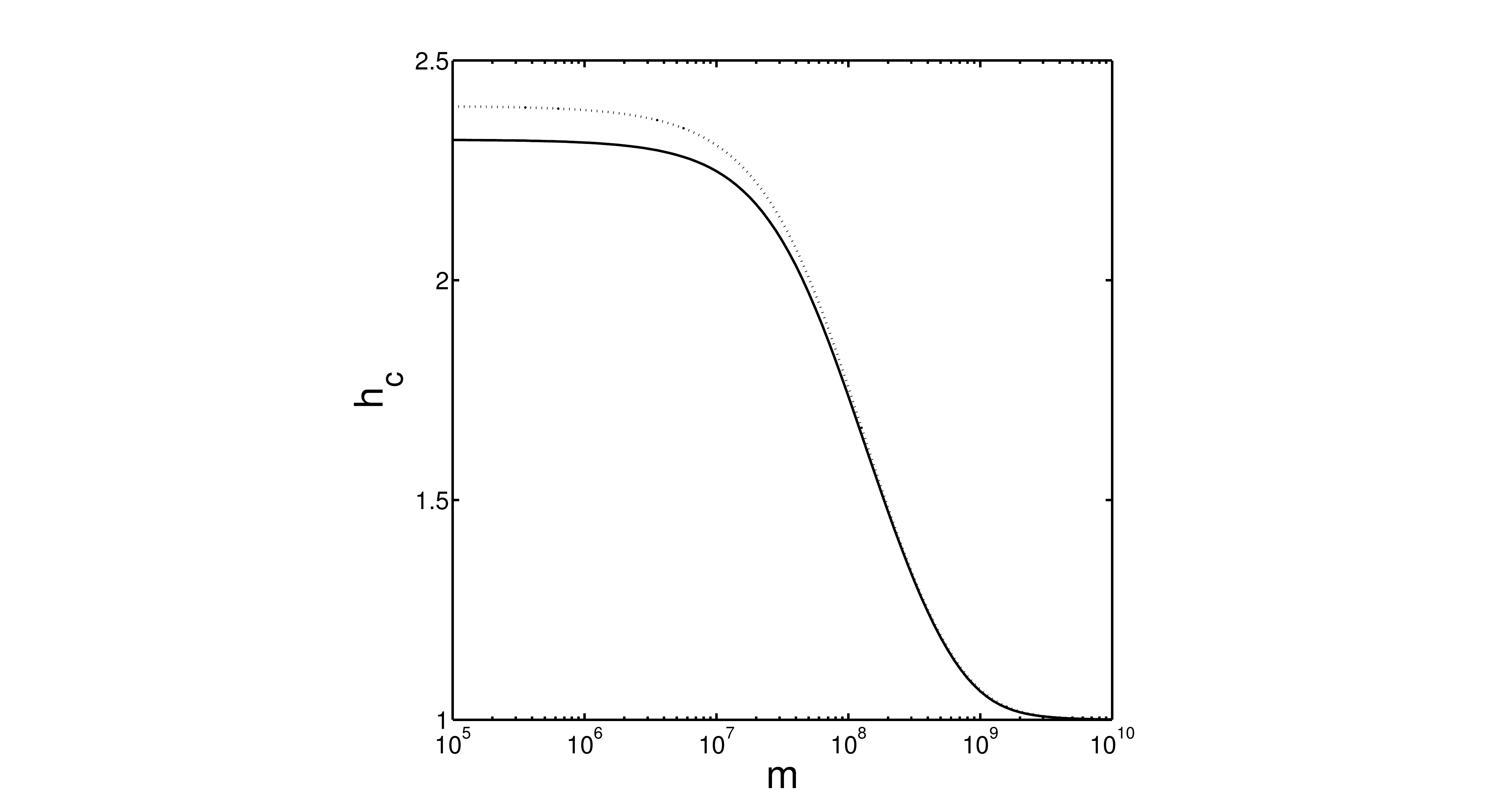}
\caption{Plot of $h_c$ as a function of the mass for $s=0.2$ and  $\xi = 60$. The solid line is the result of the numerical analysis, the dotted one is obtained with the analytical approximation described in Sec.\ \ref{sec6}.}
\centering
\label{hc_vs_m2}
\end{center}
\end{figure}
The numerical relation between the mass of the monopole and the value of $h_{c}$ is depicted in  Fig.\ \ref{hc_vs_m2} for a fixed compactness $s=0.2$ and a nonminimal coupling parameter $\xi=60$. The plot shows an interpolation 
between two asymptotic values at small and large masses. For large masses, the value of $h_c$ is bounded from above by 
$h^{\rm in}_{\rm eq}\,$ in Eq. (\ref{h_eq_in})
which converges to $h^{\rm in}_{\rm eq}=1$ for $m\approx 10^9 \rm \,\,kg$ (with $s=0.2$
and $\xi=60$). At small masses, the central value $h_c$ is independent of the mass because the Higgs potential 
contributes very little to the effective potential inside the matter distribution (see also Fig. \ref{dVeffdh}). 
\begin{figure}
\begin{center}
\includegraphics[scale=0.37,trim=270 0 310 0,clip=true]{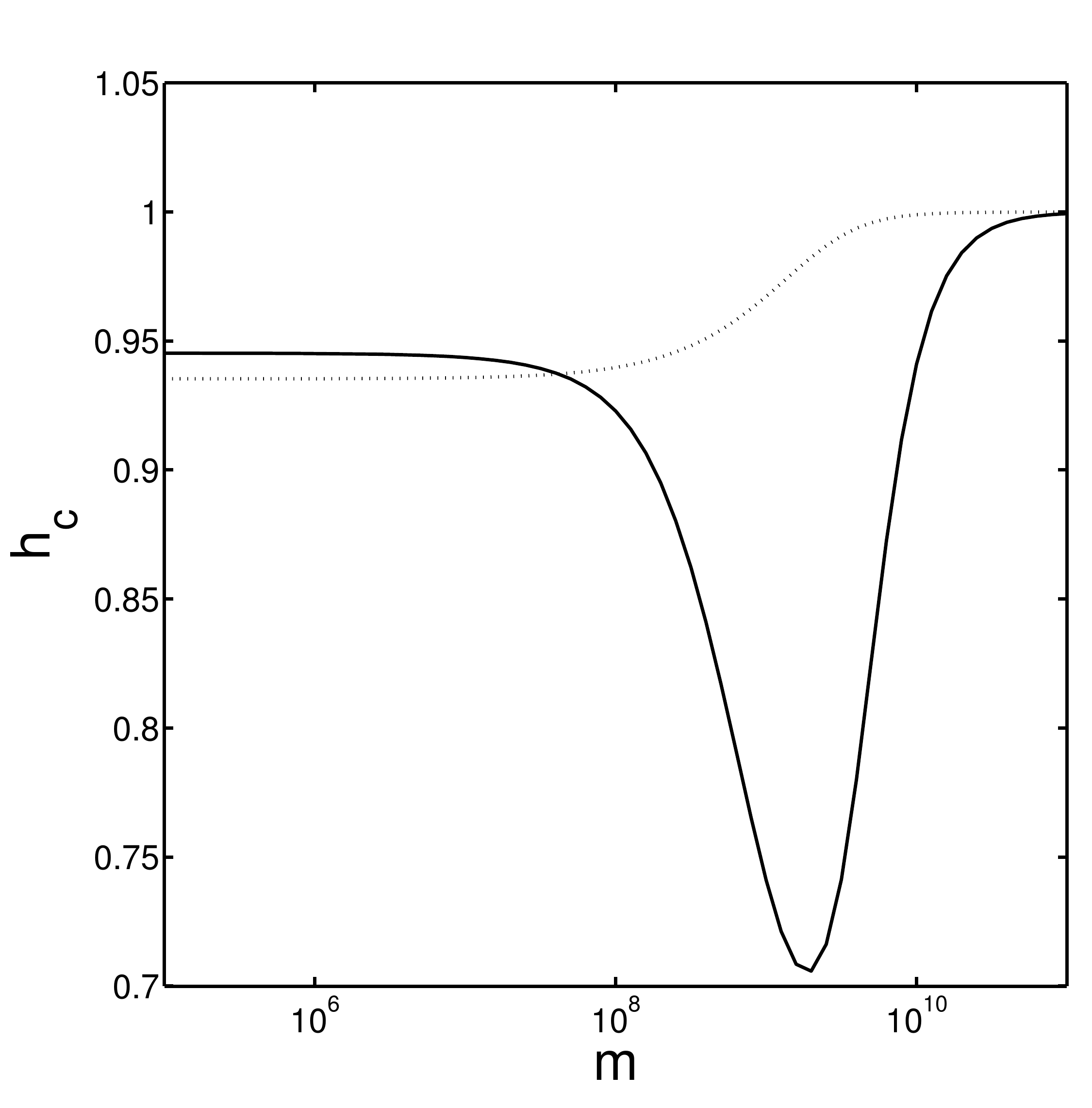}
\caption{Plot of $h_c$  in function of the mass 
for  $s=0.73$ and  $\xi = 60$. The solid line is the result of the numerical analysis while the dotted one is obtained with the analytical approximation described in Sec.\ \ref{sec6}. We see that the analytical approximation does not work well with large $s$.}
\centering
\label{hc_vs_m1}
\end{center}
\end{figure}
In Fig.\ \ref{hc_vs_m1} we show that this behavior is present also for large compactness ($s=0.73$), which 
yields  $|h_c|<1$, as seen above.
In Fig.\ \ref{dVeffdh} we represent the derivative of the effective potential $V_{\rm eff}$ given in Eq.\ \eqref{Veff} inside the matter 
distribution as a function of the mass of the monopole for fixed $\xi$ and $s$. Local maxima and minima, where $dV_{\rm eff}/dh=0$, are marked by the peaks appearing on the plot. We see that $h=0$ is always a minimum while there are two maxima at $h^{\rm in}_{\rm eq}$ 
(see Eq.(\ref{h_eq_in})), whose value converges to one for large masses. From the expression of the effective potential \eqref{Veff} (with averaged Ricci scalar)
\bea\label{Veffav}
V_{\rm eff}=-V+\frac{\xi H^2\langle R\rangle}{16\pi},
\eea
and the behaviour of $\langle R\rangle$ (see fig.\ \ref{Rmean}) we deduce that the  term $\xi H^{2}\langle R\rangle/(16\pi)$ is dominant  for small masses and becomes negligible compared to the Higgs 
potential for large masses. Thus, for small masses, the field behaves inside the matter distribution as if there was no potential, in a way 
similar to that in spontaneous scalarization \cite{sudarsky}. This is not the case, however, outside the body because here the Higgs potential can no longer be neglected compared to the nonminimal coupling term.

\begin{figure}[h]
\begin{center}
\includegraphics[scale=0.39,trim=300 0 0 0,clip=true]{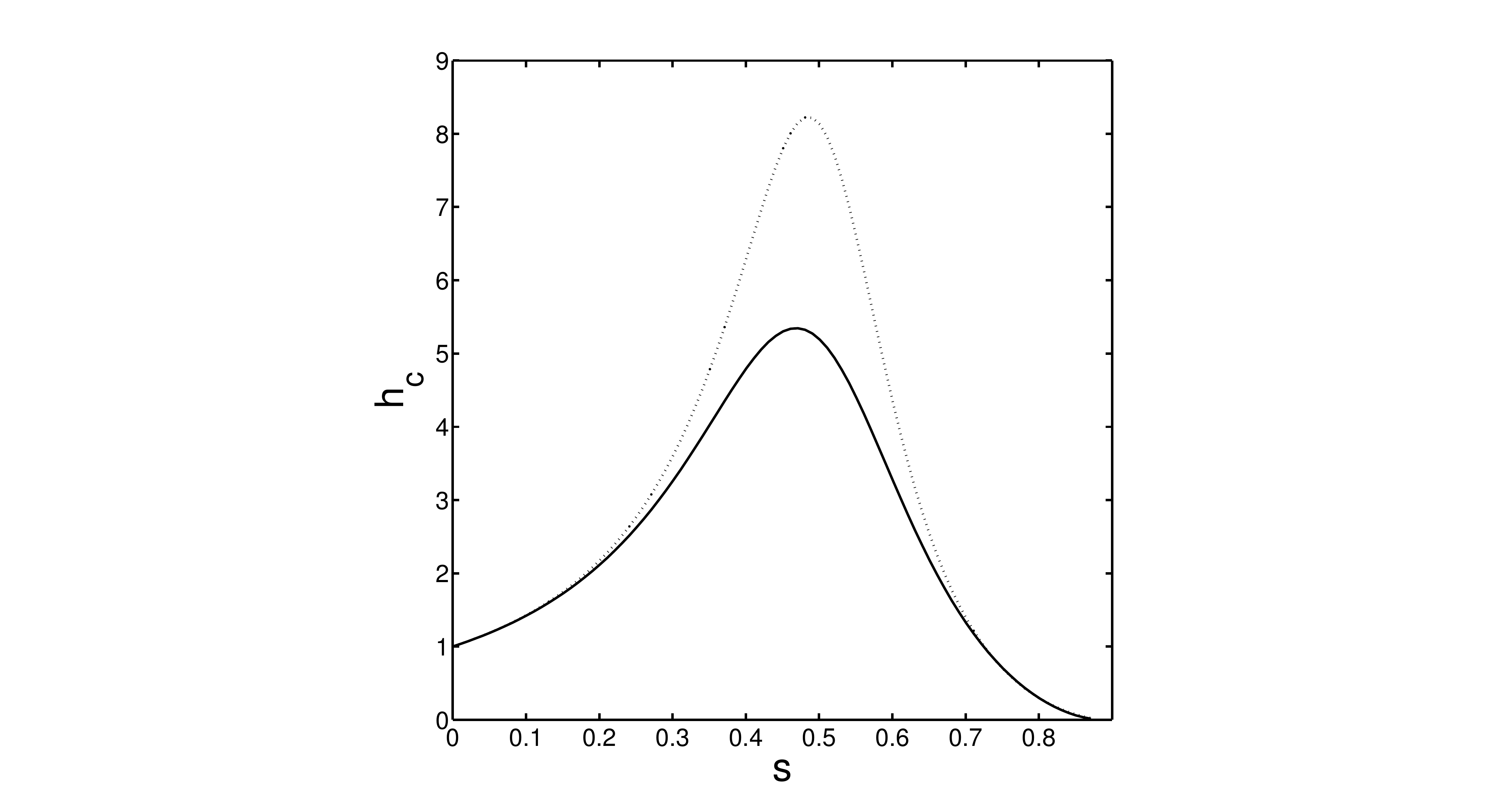}
\caption{
The plot of $h_c$ shows a peak at some value of  $s$ and for fixed $\xi$. In the case plotted here (solid line) we choose $m=100\rm \;kg$ and $\xi = 55$. The dotted line is obtained with the analytical approximation described in Sec.\ \ref{sec6}.}
\centering
\label{hc_vs_s}
\end{center}
\end{figure}

What fixes the central value $h_c$ of the monopoles is a non-linear phenomenon  of classical resonance. In
Fig.\ \ref{hc_vs_s} we show an example, where $h_c$ increases around a specific value of the compactness.  For small values of $s$, $h_c$ is close to one and the monopole distribution is pretty close to the 
homogeneous GR solution $h(r)=1$. We find that, for astrophysical objects like the Sun, the 
combination of very low compactness and large mass  makes the Higgs field extremely close to its vev everywhere, yielding negligible deviations from GR. This is in line with the PPN analysis presented in Sec.\ \ref{PPN}.  On the other hand, we have seen that, for $s>0.7$, $|h_c|$  is smaller than one, since $\langle R\rangle$ is negative. Between these two extreme cases, there exists a specific 
value of $s$ that maximizes $h_c$. This is a new result due to the combined action of the nonminimal coupling and the field potential. In fact, it is absent if the potential vanishes as in \cite{sudarsky}. 

\section{Amplification mechanism}
\label{sec6}

\noindent In this section, we present an analytical model of the scalar field resonant amplification found numerically in the previous section. As before, we consider the Klein-Gordon equation Eq.\ \eqref{kg_simp} in dimensionless units and
we suppose that the metric fields $\lambda$ and $\nu$ are the ones given by GR. The combination of Eqs.\ \eqref{nu} and \eqref{lam} gives
\bea
\nu_u-\lambda_u+\frac{2}{u}=
 \left\{\begin{array}{ll} 
	  -{{uR}\over {6\left(1-s^3u^2\right)}}+\frac{2}{u}  & 0<u<1/s,
	  \\{{2u-1}\over{u\left(u-1\right)}}\approx {2\over{u-1}} & u>1/s,\end{array}\right.
\eea
where the top (bottom) line corresponds to the internal (external) solution.

We now expand $dV/dh$ around  $h=h_*$, where $h_*=1$ outside and $h_*=h_c$ inside the body. The function  to be expanded has the form
\bea
f(h)=\alpha h \left(h^2-1\right),
\eea
where $\alpha=2 \lambda_{\rm sm} r_s^2 m_{pl}^2\tilde v^2$, thus, up to the first order, we have
\bea
f(h)\approx \alpha \left[ h_* \left(h_*^2-1\right)+\left(3h_*^2-1\right)\left(h-h_*\right) \right].
\eea
We now examine more carefully the external and internal solutions of the Klein-Gordon equation.

\subsection{External solution}

\noindent For the external solution, we can assume that $u\gg 1$, $R\simeq0$ (like for the numerical treatment), $h\simeq1$ since the Higgs field settles to its vev at large distance, $\nu_u-\lambda_u+\frac{2}{u}\simeq \frac{2}{u}$, and $e^{2\lambda}\sim1$. After the expansion, upon the change of variables $h=Y/u$ and $Z=Y-u$,  Eq.\ \eqref{kg_simp} can be written as 
\bea
Z_{uu}=\alpha Z,
\eea
which has the general solution
\bea
h_{\rm ext}={\mathcal{C}_1\over u} e^{\sqrt{\alpha}u}+{\mathcal{C}_2\over u} e^{-\sqrt{\alpha}u}+1,
\label{h_ext}
\eea
for arbitrary constant $\mathcal{C}_1$ and $\mathcal{C}_2$. The requirement  $\lim_{u\rightarrow \infty}h=1$ yields $\mathcal{C}_1=0$, so we find the Yukawa distribution for the Higgs field outside the compact object given by
\bea
h_{\rm ext}= \frac{Q}{u} e^{-u/L}+1.
\eea
The parameters $Q=\mathcal{C}_2$ and $L=1/\sqrt{\alpha}$ can be identified as the  scalar charge and the characteristic length respectively, which justifies the term ``monopole'' used to name these solutions.  To fix $\mathcal{C}_2$, we will use the continuity condition of the Higgs field at the boundary of the compact object given by $h_{\rm ext}(1/s)=h_{\rm int}(1/s)$. In addition,  the continuity condition of the derivative, $h'_{\rm ext}(1/s)=h'_{\rm int}(1/s)$, will lead to an implicit equation for $h_c$.

\subsection{Internal solution}

\noindent We now derive the analytical Higgs field profile for the internal region. We make the same assumption as before for the terms involving $\nu$ and $\lambda$, but we assume $u\simeq0$ and $R\sim \langle R\rangle \neq0$.
We now expand $f(h)$ around $h_*=h_c$ and change the variables according to $h=Y/u$ and
\bea
Z=Y+\frac{B(h_c)}{A(h_c)} u,
\eea
where 
\bea
A(h_c)&=&\frac{\alpha}{2} \left(3h_c^2-1\right) - {{ \langle R\rangle\xi}\over{8\pi}}, 
\label{const_A} \\
B(h_c)&=&-\alpha h_c^3.
\label{const_B}
\eea
We then obtain the differential equation 
\bea
Z_{uu}=A\left(h_c\right) Z,
\eea
for which it is sufficient to discuss the solution for $A\left(h_c\right)<0$, the positive case being basically the same. The case $A\left(h_c\right)=0$   is not considered as it corresponds to an uninteresting fine-tuning of the parameters. The solution reads
\bea
h_{\rm int}=\frac{\mathcal{D}_1}{u} e^{\sqrt{A}u}+\frac{\mathcal{D}_2}{u} e^{-\sqrt{A}u}-\frac{B}{A},
\eea
where $\mathcal{D}_1$ and $\mathcal{D}_2$ are constants of integration. The condition of regularity of the Higgs field at the origin, $h_{\rm int}(u=0)=h_c$ implies that $\mathcal{D}_1=-\mathcal{D}_2$. In addition, when $u\rightarrow0$ we also find that
\bea
\mathcal{D}_1=\frac{1}{\sqrt{|A|}} \left(h_c+\frac{B}{A} \right),
\label{const_int}
\eea
so, the linearized expression for the Higgs field inside the compact object is given by
\bea
h_{\rm int}=\frac{\mathcal{D}_1}{u} \sin \left(\sqrt{|A|}u\right)-\frac{B}{A}.
\label{h_int}
\eea
As mentioned above, the conditions of continuity of the Higgs field and its derivative allow us to fix  $\mathcal{C}_2$
and to derive an implicit equation for determining $h_c$. Indeed, by imposing $h_{\rm ext}\left(1/s\right)=h_{\rm int}\left(1/s\right)$, we find that
\bea
\mathcal{C}_2=\frac{1}{s} e^{\frac{\sqrt{\alpha}}{s}} \left(\mathcal{D}_1 s \sin\left(\sqrt{|A|}/s\right)-\frac{B}{A}-1 \right),
\label{const_ext}
\eea
while the regularity condition $h'_{\rm ext}\left(1/s\right)=h'_{\rm int}\left(1/s\right)$ yields the implicit equation
\bea\non
&&\left(h_{c}+{B\over A}\right)\left[ \sqrt{\alpha\over |A|}\sin \left(\sqrt{|A|}\over s\right)+\cos \left(\sqrt{|A|}\over s\right)   \right]\\
&&\hspace{2.5cm}=\left(1+{B\over A}\right)\left(1+{\sqrt{\alpha}\over s}\right).
\label{simplifed_Ahc}
\eea
The solution for the case $A\left(h_c\right)>0$ can be found by replacing the sine and cosine by hyperbolic sine and hyperbolic cosine. 
However, the condition $A\left(h_c\right)<0$ is  necessary  for the resonant amplification.
The expression \eqref{simplifed_Ahc} greatly simplifies when $\alpha$ is small as for macroscopic bodies \footnote{As an example, for an object of the mass range of an asteroid ($M\simeq 10^{7}$), $\alpha\simeq 10^
{-25}$.}. In fact,  since $B/A\simeq 0$ when $\alpha$ is negligible, the implicit equation for $h_c$ Eq.\eqref{simplifed_Ahc} reduces to 
\bea
h_{c}=\left| \cos\sqrt{{\xi \langle R\rangle}\over {8\pi s^2}}\right|^{-1}
\label{approx_alpha_small}
\eea
where the absolute value is necessary when the positive $h_{c}$ branch is chosen. We see that, in this approximation, the central value of the Higgs field $h_c$ has periodic divergences corresponding to certain values of $s$, $\xi$, and of the mass of the monopole.
E.g. for asteroids, the compactness is very small ($s\sim 10^{-12}$) and one finds that $h_{c}=1$ to great accuracy. Notice that, for small $s$, the condition $A<0$ is no 
longer true and we must switch $\cos$ to $\cosh$, which yields, however, the same result. We thus confirm the results obtained in the previous section:  for small values of the compactness, the central value of the Higgs field $h_c$ is very close to the Higgs vev. 
For larger values of the compactness, the approximate formula \eqref{approx_alpha_small} shows that, for a given monopole mass and  $\xi$,  $h_c$ has peaks corresponding to critical values of $s$. These are the resonances that we have also seen numerically. The number of peaks depends on the nonminimal coupling $\xi$ as we will see on the next section.
Note that the condition $A\left(h_c\right)<0$ is favored by a large nonminimal coupling (see Eq.\ \eqref{const_A}), and so the approximate equation \eqref{approx_alpha_small}, is even more accurate in the large $s$ regime. As we will see in the next section, there exists a critical value of $\xi$ for which one peak splits into two separate peaks. Another interesting limit is $\sqrt{A}/s\ll 1$. In this case, the formula reduces to
\bea
\left(h_{c}+{B\over A}\right)\left(1+{\sqrt{\alpha}\over s}\right)
\simeq\left(1+{B\over A}\right)\left(1+{\sqrt{\alpha}\over s}\right)
\eea
which implies that $h_{c}\sim 1$. 
The regime $A/s^{2}\sim 0$ corresponds to  
\bea
r_{s}={16\pi\lambda v^{2} \mathcal{R}^{3}\over 3\xi}
\eea
where $\mathcal{R}$ is the radius of the compact object (assuming $\langle R\rangle\approx 3 s^3/r_s^2$). This relation can be written again as $s\xi\simeq(10^{18}\mathcal{R})^{2}$ with $\mathcal{R}$ expressed in meters. 
It is then obvious that this regime is totally unphysical unless $\xi$ is very large \footnote{A very large $\xi$ is not excluded by LHC experiments, see \cite{calmet}.}.

\subsection{Analysis of the parameter space}

\noindent 
As we saw in Sec.\ \ref{sec5},  Figs.\ \ref{hc_vs_m2}, \ref{hc_vs_m1}, and \ref{hc_vs_s}, 
the value of $h_c$ in function of the parameters can be qualitatively reproduced with our analytical model with good accuracy. There are some discrepancies (see for instance Fig.\ \ref{hc_vs_m1} for large compactness) but the analytical model expressed by Eq.\ \eqref{simplifed_Ahc} Êis sufficient to understand the amplification mechanism. For example, in Fig.\   \ref{hc_vs_s} we see a good agreement between our analytical model and the full solution for the position of the resonance, although there is some overestimation of 
its amplitude.

In the rest of this section, we will use the analytical model to explore the parameter space of the  monopole, given by mass, compactness, and nonminimal coupling. Once these  are fixed,  the 
central value of the Higgs field is uniquely determined by the implicit equation \eqref{simplifed_Ahc}. 

\begin{figure}
\begin{center}
\includegraphics[scale=0.32, trim=380 0 0 0,clip=true]{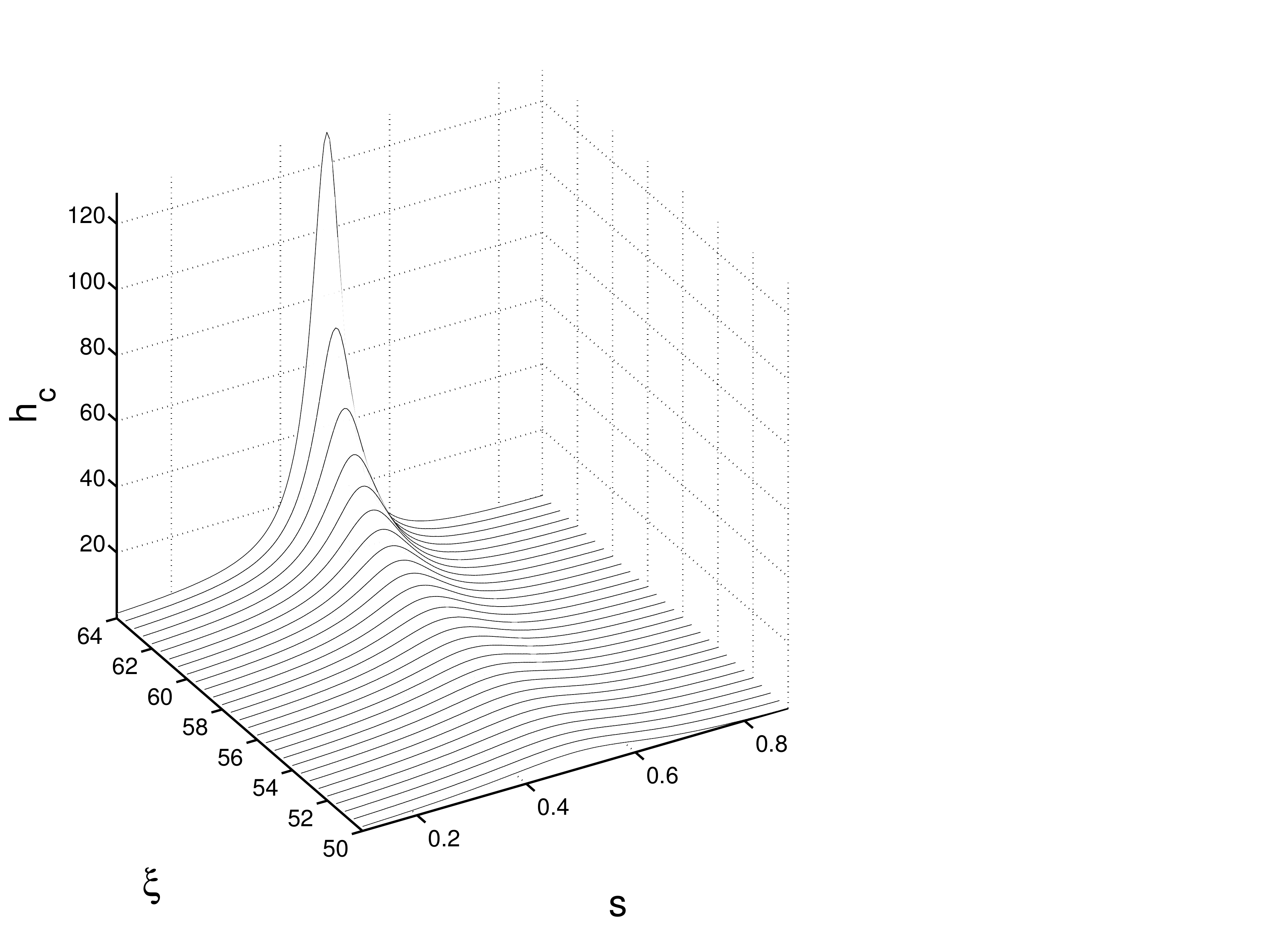}
\caption{Plot of $h_c$ in function of $s$ and $\xi$ as given by the implicit relation \eqref{simplifed_Ahc} for a fixed mass $m=10^3\rm kg$.
We see that the peak sharpens for increasing $\xi$.}
\centering
\label{hc_vs_sxi1}
\end{center}
\end{figure}

\begin{figure}
\begin{center}
\includegraphics[scale=0.30,trim=325 0 0 0,clip=true]{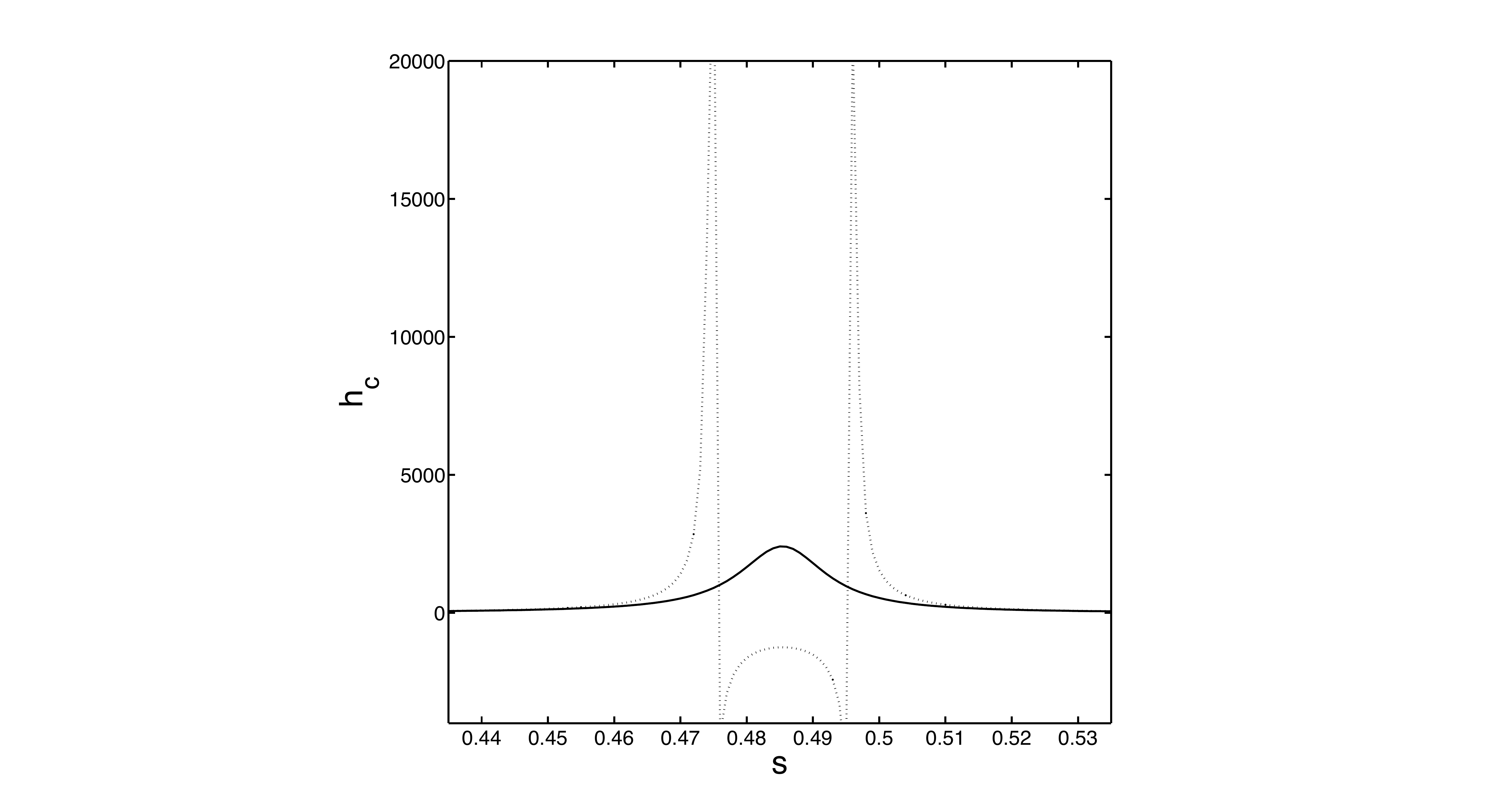}
\caption{Plot of $h_c$ given by the implicit Eq.\ \eqref{simplifed_Ahc}
in function of the compactness for $\xi=64.6$ (solid line) and $\xi=64.7$
(dashed line). The monopole mass is fixed at $m=10^3\rm kg$.}
\label{hc_vs_sxi2}
\end{center}
\end{figure}

In Fig.\ \ref{hc_vs_sxi1} we show how the resonance in $h_c$ evolves as a function of the compactness and of the nonminimal coupling. 
As $\xi$ increases, the peak grows and sharpens. The question is then how large the resonance can be. It seems that there exists a critical value of $\xi=\xi_{\rm cr}$ above which  $h_c$ diverges. This is illustrated in Fig.\ \ref{hc_vs_sxi2} where we plotted $h_c$ 
for  both $\xi<\xi_{\rm cr}$ and $\xi=\xi_{\rm cr}$. The two vertical asymptotes in $h_c$ appear when the nonminimal coupling becomes larger than $\xi=\xi_{\rm cr}$ and they correspond to a phase transition, in which $h_c$ switches sign. We recall in fact that there are two branches corresponding to $v=\pm 246$ GeV. Even though we chose $v$ to be the positive root, there is still the possibility that $h(r)$ jumps to the negative branch, which is a perfectly valid mathematical solution of the Klein-Gordon equation \footnote{This problem could be avoided by considering a Higgs multiplet with an Abelian $U(1)$ symmetry. The amplitude and the phase of the Higgs field would be under a much better analytical and numerical control.}.

\begin{figure}
\begin{center}
\includegraphics[scale=0.31,trim=350 0 380 0,clip=true]{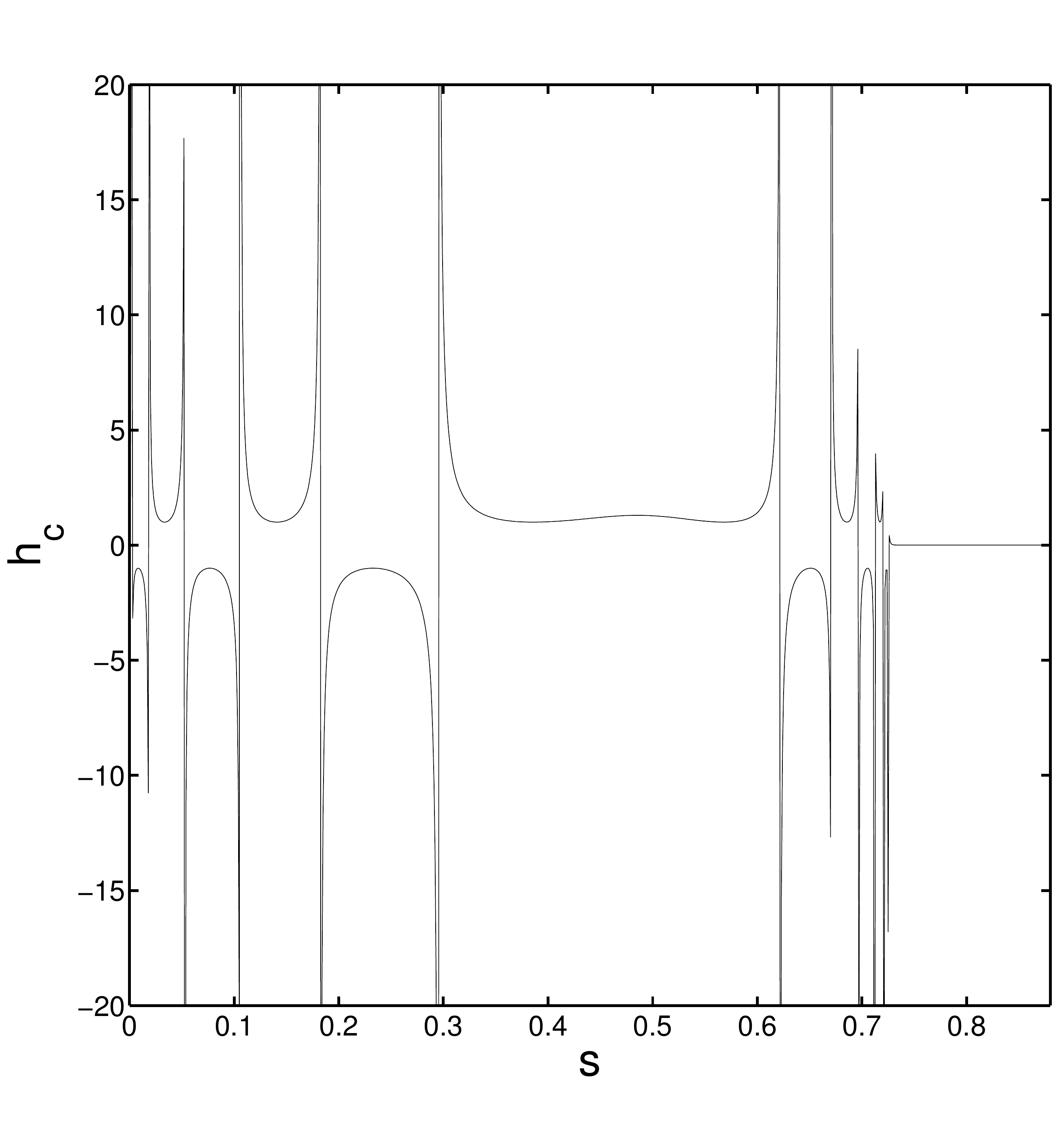}
\caption{Plot of $h_c$ in function of $s$ for $\xi=10^{4}$ and $m=10^2\rm kg$ obtained from the expression \eqref{simplifed_Ahc}.}
\centering
\label{hc_vs_sxi1e4}
\end{center}
\end{figure}

\begin{figure*}[ht]
\begin{center}
\begin{tabular}{ccc}
\includegraphics[scale=0.2,trim=415 0 240 0,clip=true]{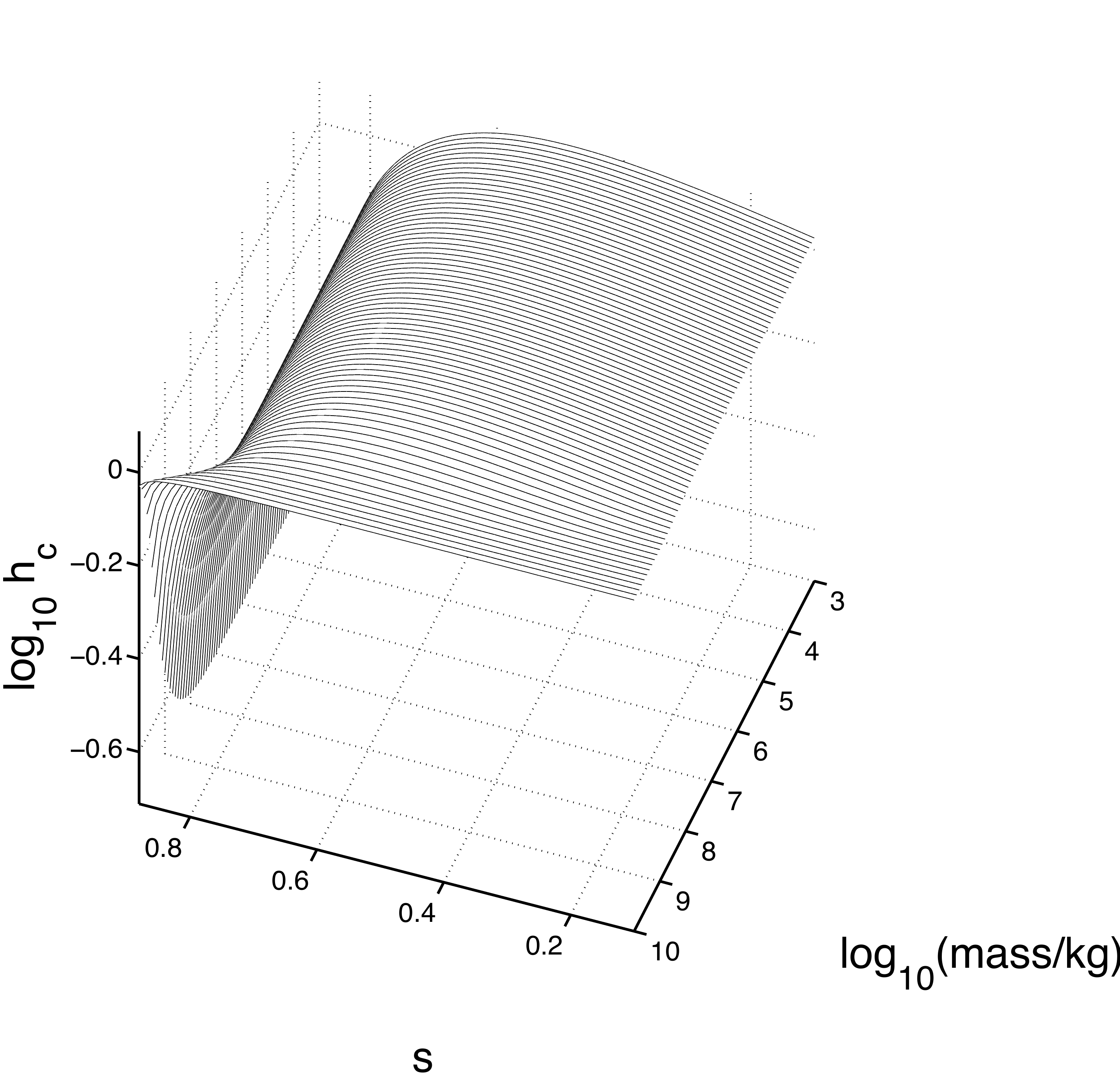} &
\includegraphics[scale=0.2,trim=415 0 240 0,clip=true]{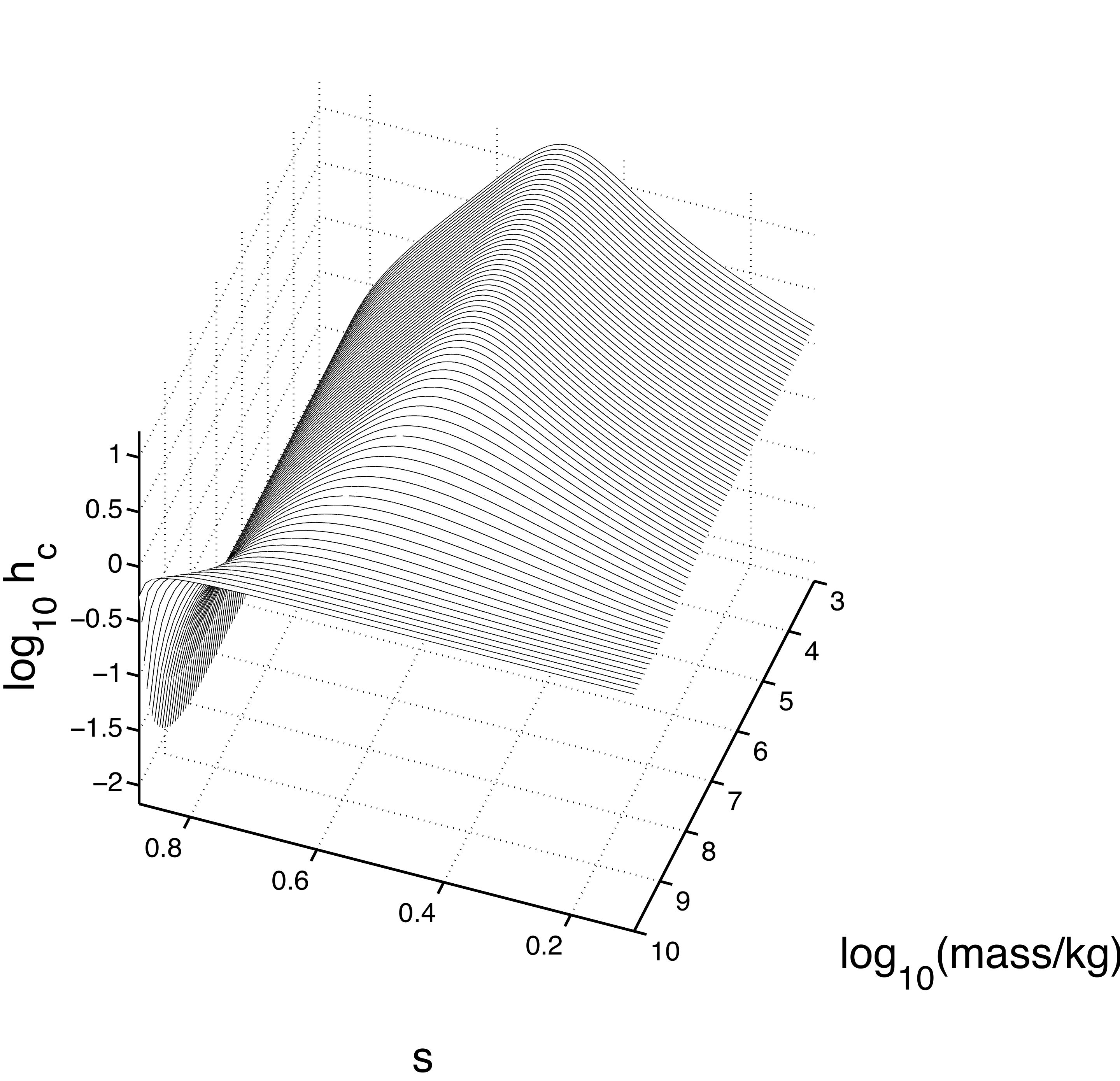} &
\includegraphics[scale=0.2,trim=415 0 240 0,clip=true]{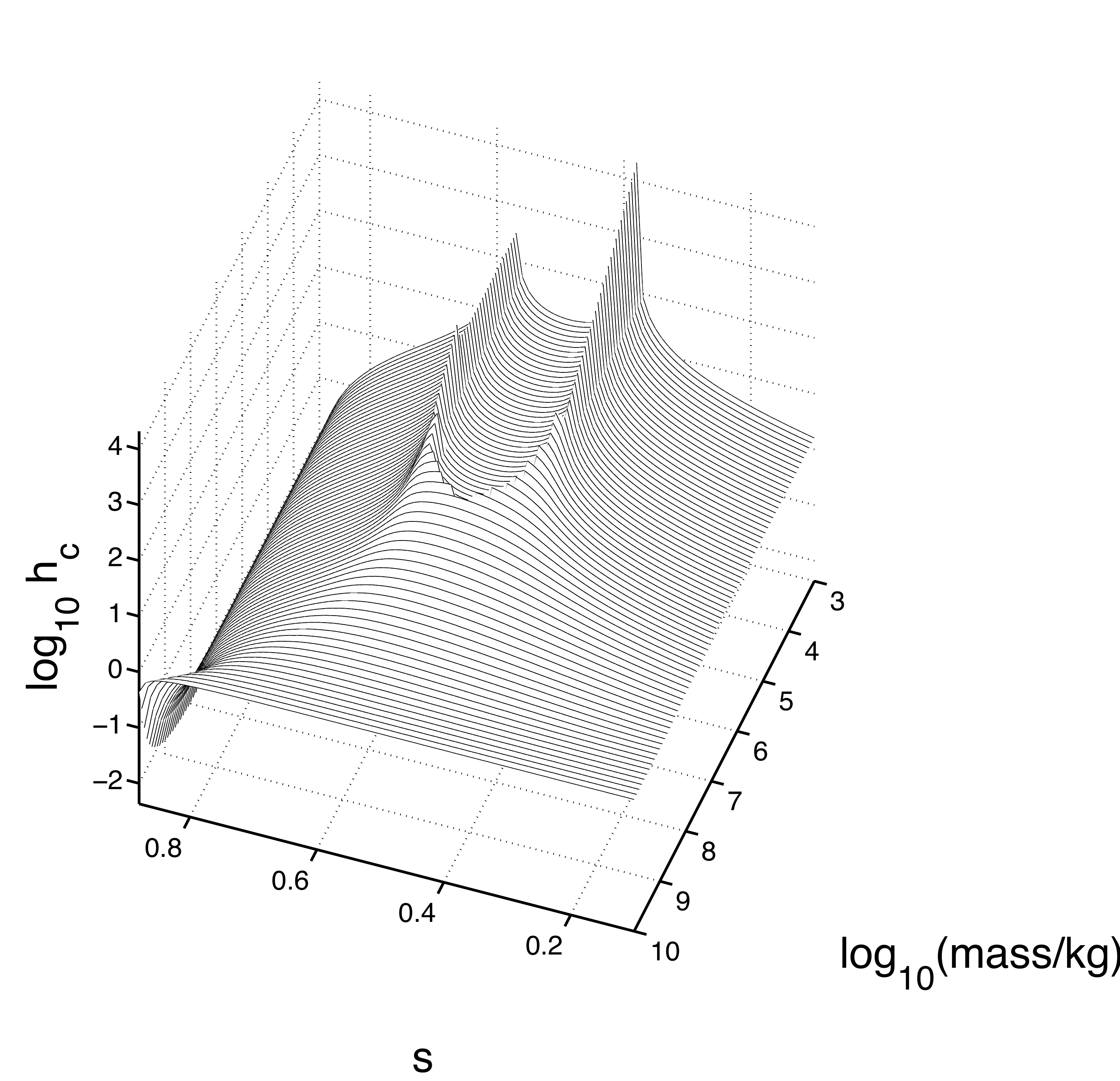} 
\end{tabular}
\end{center}      
\caption{Plot of $h_c$ (obtained with the analytical approximation) in function of the mass of the monopole and its compactness for $\xi=10, 60, 70$ (from left to right).}                    
\label{hc_ana_ifo_xi}  
\end{figure*}   

This also implies that, when the nonminimal coupling is larger than  $\xi_{\rm cr}$, there can be forbidden values for the compactness (or, equivalently, for the size) in the parameter space. As an example, we plot in Fig.\ \ref{hc_vs_sxi1e4} $h_c$ in function of the compactness for $m=10^2$ kg and $\xi=10^4$, which corresponds to the value predicted by Higgs inflation \cite{shaposh}. We see that there are multiple divergences, also for relatively small values of $s$.
However, this does not prevent the nonminimal coupling parameter to be arbitrarily large since   $\xi_{\rm cr}$ basically depends on the mass of the monopole.


Let us look at Fig.\ \ref{hc_ana_ifo_xi}, where we plot  $h_c$ in function of the monopole mass and  compactness for three values of 
$\xi$. We see that $h_c$ generically settles to its vev,  $h_c=1$ for small compactness and large mass. The peaks appear for small masses (see the right panel of fig.\ \ref{hc_ana_ifo_xi}) and sharpen as $\xi$  increases, until $h_{c}$ eventually diverge at some $\xi_{\rm cr}$. In the small mass regime the Higgs potential is much smaller than the coupling term, which is proportional to $\xi$, see Eq.\ \eqref{Veffav}. In the large mass regime, however, the upper bound imposed by Eq.\ \eqref{h_eq_in} becomes closer and closer to one. These two competing effects explain qualitatively the presence of the peaks in the small mass region rather than in the large mass one, provided the compactness $s$ is not too small, otherwise $\langle R \rangle $ is too small and always smaller than the Higgs potential. In such a case, scalar amplification is negligible, no matter the monopole mass. In summary, for large values  of the nonminimal coupling, monopoles with small masses cannot exist for certain values of the compactness for which the Higgs field at the center of the body diverges. On the opposite, large mass monopoles always exist but the scalar amplification is much smaller.

\section{Conclusion}\label{conclusions}

\noindent In this work we have studied in detail the static and spherically symmetric solutions, originally found in \cite{monopole}, of the equations of motion of a theory where the Higgs field in unitary gauge is nonminimally coupled to gravity. This model is inspired by Higgs inflation, where the coupling strength $\xi$ is very large. We have thus verified that such a large coupling is not in conflict with experiments when we look at the same theory at energy densities typical of compact objects such as the Sun. However, the main result of this paper is the existence of particlelike solutions that are asymptotically flat and have finite classical energy. In particular, we have found a new non-linear mechanism of resonant amplification that is not present in the models with vanishing potentials studied so far. We explored this amplification mechanism numerically and found an analytical approximation that shows that, in the large coupling regime, there are forbidden combinations of radius and baryonic density, at which the value of the Higgs field at the center of the spherical body tends to diverge. As for spontaneous scalarization, the amplification mechanism found here is a general feature that can be applied to cases with different parameters and/or potential shapes.

In principle, the shift in the Higgs expectation value inside the compact object leads to a change in the mass of the $W$ and $Z$ bosons that, in turn, has an impact on the mass of decay products, decay rates and so on. In the case at hand, however, we have seen that the shift in objects like neutron stars is negligible. Larger effects are possible only in ranges of mass and density that are very unphysical, as shown in Table I. Therefore, in realistic compact objects we do not expect any observable modification. Similarly, in the case of a Yukawa coupling to fermions we do not expect any dramatic effect for the same reasons. Technically, the addition of a Yukawa coupling to fermions would introduce in the Klein-Gordon equation new terms (one for each fermion) proportional to $H$, which will compete in the dynamics with the non-minimal coupling term $\xi R$, see Eq.\ \eqref{KG}. However, for realistic objects, this contribution will be much larger than the gravitational one so we do not expect significant deviations from the case with ordinary equations of state. We point out that this is a very different situation as in previous models, like \cite{sudarsky}, where the amount of spontaneous scalarization was much larger. In our case,  the presence of Higgs potential in the action effectively prevents fundamental interactions to change inside compact objects. 

From the quantum field theory point of view, a r-dependent vacuum leads to a non-local effective action whose effect are very small in the regime of small curvature considered here but might be important in the primordial Universe or in strong field configurations \cite{shapiro}. 

We remark that the top-hat profile \eqref{rho} is a simplifying hypothesis that saves some computational effort. Introducing a different profile, and a specific equation of state, would be more realistic but no substantial changes are expected in our results. This claim is supported by some previous work (e.g. \cite{sudarsky,esposito}) and by some numerical tests that we performed by smoothing out the step function.

About the stability of our solutions, we point out that any change of the value of the Higgs at the center of the body leads to a change in the geometry of the spacetime at infinity. The solutions that we have found are asymptotically flat while any other solution is asymptotically de Sitter, as discussed in \ref{sec3}. In a realistic scenario of a spherical collapse in an asymptotically flat spacetime we expect that the Higgs monopoles are the only solutions and are stable. A formal proof of this statement would require the study of perturbations, along the lines of \cite{Volkov} around the (numerical) monopole solution and goes beyond the scope of this paper, although it is a very interesting question.

There are several aspects that deserve further analysis. For instance, in this paper we assumed that the characteristic parameters are the ones of the standard model (in particular the coupling $\lambda_{\rm sm}$ and the vev $v$, see \eqref{mexican}). As a result, the deviations from GR are negligible. It would be interesting to find to what extent these parameters can vary without violations of the current observational constraints.  Finally, we believe that also the symmetry structure of the Higgs field and its influence on the solutions should be studied. In fact, the Higgs field should be treated as a complex multiplet with $SU(2)$ gauge symmetry and it is know that the dynamical effect of the Goldstone components might play a role in inflation \cite{kaiser} but also at low energy \cite{maxhiggs}. In relation to this, we also recall that there exist exact solutions for Abelian and non-Abelian configuration in Minkowski space called Q-balls \cite{Qballs}. In the baryonic massless limit, but with the gauge symmetry restored, Higgs monopoles could be generalized to describe gauged Q-balls in curved space. We feel that this direction should be explored as it might lead to discover solutions with physical properties that are compatible with dark matter. If not, it would nevertheless interesting to see if these solutions are excluded by precise Solar System tests.

\begin{acknowledgments}
\noindent All computations were performed at the ``plate-forme technologique en calcul intensif'' (PTCI) of the University of Namur, Belgium,
with the financial support of the F.R.S.-FNRS (convention No. 2.4617.07. and 2.5020.11). M.R.\ is partially supported by the ARC convention No.11/15-040 and S.S.\ is a FRIA Research Fellow.
\end{acknowledgments}


\appendix

\section{Alternative weak-field analysis}\label{appendix1}

\noindent We follow here the discussion on the weak field approximation presented in \cite{esposito}. We first write the action of the theory in Einstein frame, 
starting from the Lagrangian Eq.\ \eqref{BDaction}. With the conformal transformation $g^{*}_{\mu\nu}=\phi g_{\mu\nu}$ and the field redefinition
$2 d \varphi=\sqrt{2\omega +3}d \ln \phi$, we find the canonical Einstein action action  
\bea\non
S&=&{c^{3}\over 4\pi G^{*}}\int d^{4}x\sqrt{g^{*}}\left[{R^{*}\over 4}-{1\over 2}g^{\mu\nu}_{*}\partial_{\mu}\varphi\partial_{\nu}\varphi-U(\varphi)\right]\\
&+&S_{m}[\Psi,\phi^{-1}g_{\mu\nu}^{*}],
\eea
where $G^{*}$ is the Newton constant in Einstein frame, $U(\varphi)=\bar V/\phi^{2}$, and where we reintroduced the speed of light $c$ to perform the correct expansion of the field equations. The Klein-Gordon equation reads
\bea
\square^{*}\varphi=-{4\pi G^{*}\over c^{2}}\alpha(\varphi)T+{dU\over d\varphi},
\eea
where we defined the matter scalar coupling $\alpha^{2}=(2\omega+3)^{-1}$. By expanding the field in power of $c^{-2}$ we find
\bea\label{KGapp}
\Delta \varphi \simeq {4\pi G^{*}\over c^{2}}\alpha_{0}\rho+m^{2}_{\varphi}\varphi+{\cal O}\left(1\over c^{4}\right),
\eea
where $\Delta$ is the flat Laplacian. To obtain this equation, we expanded $\alpha=\alpha_{0}+\alpha_{1}\varphi+\ldots$, the trace of the matter stress tensor 
$T=-\rho c^{2}+\ldots$ and we wrote $dU/d\varphi=m^{2}_{\varphi}\varphi$. The effective mass is obtained by expanding the original potential $\bar V$ around the vev
of $\phi$ and reads
\bea\non
m_{\varphi}^{2}&=&{4\lambda_{\rm sm}\pi m_{p}^{4}\over \xi^{2}(m_{p}^{2}+\xi v^{2})}\sqrt{3+{4\pi\over \xi}+{4\pi m_{p}^{2}\over \xi^{2}v^{2}}}\\\non&\simeq& {8\lambda_{\rm sm}\pi^{3/2}m_{p}^{3}\over v\xi^{3}},\label{effmass}
\eea
where the second line follows from the fact that $\xi v^{2}/m_{p}^{2}\ll 1$. The solution to Eq.\ \eqref{KGapp} has the typical Yukawa form
\bea
\varphi(r)\sim {e^{-m_{\varphi}r}\over r},
\eea
from which one can see the distance scale over which the field is significant. In our case, for $\xi$ of the order of $10^{4}$, we find from Eq.\ \eqref{effmass}
that $m_{\varphi}\simeq 4\times 10^{21}\,\, {\rm GeV}$  so the range is negligible. This analysis is valid only in the linear regime and we have proven numerically that
there exist resonances for which the scalarization is strong and the range is much larger. For these, the weak-field analysis shown here is not valid.



\section{Numerical methods}\label{appendix2}

\noindent In Sec.\ref{sec5} we present monopole solutions obtained by a simplified integration of the Klein-Gordon equation only. To show that this method is accurate, here we present the results of the integration of the full set of equations of motion, namely the Klein-Gordon together with the Einstein equations \eqref{eom_tensor}. This result confirms  that we can safely replace the metric inside the compact object by the standard GR metric as explained in Sec.\ \ref{sec3}.

\subsection{Equations of motion}

\noindent We first list the full set of equations to be solved. The explicit  $tt-$, $\theta\theta-$ and $rr-$ components of the Einstein equations \eqref{eom_tensor} are, respectively
\begin{widetext}
\bea
\nu''+\nu'^{2}-\lambda'\nu'+(\nu'-\lambda')\left({1\over r}+{H'\over F }{dF\over dH}  \right)+{1\over F}{dF\over 
dH}\left(H''+{H'\over r}\right)+{H'^{2}\over F}\left({\kappa\over 2}+{d^{2}F\over dH^{2}}\right)+\left(\kappa 
V-{3p\over \mathfrak{R}^{2}\rho}\right){e^{2\lambda}\over F}=0,
\label{eom_r}
\eea
\bea
\lambda'\left({2\over r}+{H'\over F}{dF\over dH}\right)-{H''\over F}{dF\over dH}-{H'^{2}\over 
2F}\left(\kappa+2{d^{2}F\over dH^{2}}\right)-{2H'\over rF}{dF\over dH}-{1\over r^{2}}\left(1-
e^{2\lambda}\right)-{e^{2\lambda}\over F}\left(\kappa V+{3\over \mathfrak{R}^{2}}\right)=0,
\label{eom_t}
\eea
\bea
\nu'\left({2\over r}+{H'\over F}\frac{dF}{dH}\right)-\frac{\kappa H'^{2}}{2F}+{1\over r^{2}}\left(1-
e^{2\lambda}\right)
+{2H'\over rF}\frac{dF}{dH}-{e^{2\lambda}\over F}\left(\frac{3p}{\mathfrak{R}^{2}\rho}-\kappa V\right)=0,
\label{eom_theta}
\eea
\end{widetext}
where the prime denotes a derivative with respect to $r$ and $\mathfrak{R}^2={\mathcal{R}^3/ r_s}$, $r_{s}$ being the 
standard Schwarzschild radius. We take the energy density as in eq.\  \eqref{rho} so that $\mathfrak{R}={3\over 
\kappa\rho_0}$. Finally, the Klein-Gordon equation reads

\bea\non
H''-H'\left(\lambda'-\nu'-{2\over r} \right)+e^{2\lambda} \left({R\over 2\kappa} \frac{dF}{dH}  - \frac{dV}
{dH}\right)=0, \\
\label{KG_static}
\eea
where the Ricci scalar $R$ is given by
\bea\non
R\!=\!{2\over r^2}-e^{-2\lambda}\!\left(\!2\nu''-2\nu'\lambda'  +{4\nu'\over r}+{2\over r^{2}} -{4\lambda'\over 
r}+2\nu'^{2}\!\right)\!.\\
\label{analyticalR}
\eea

\subsection{Dimensionless system}

\noindent  To implement the numerical integration we need to write the above equations in a convenient dimensionless units system. This step is actually crucial in order to extract significant numerical results because of the involved scales, like the Planck mass. 
We first rescale the Higgs field as
\bea
H[\mathrm{GeV}]= m_p \tilde{v} h= m_p \tilde{v} \left(1+\chi\right),
\label{change_H}
\eea 
where $h$ and $\tilde{v}=246\; \mathrm{GeV}/m_p$ are dimensionless. The quantity  $\chi$ characterizes the dimensionless displacement of the Higgs scalar field around its vev. 
We express the radial coordinate in term of the standard Schwarzschild radius $r_s$ 
\be
u={r\over r_s},
\label{change_r}
\ee
and we remind that the Schwarzschild radius in Planck units is
\be
r_s [\mathrm{GeV^{-1}}]=\frac{2 m_b}{m_p^2} \times 5.61 \times 10^{26} [\mathrm{GeV\; kg^{-1}}],
\ee
where $m_b$ is the baryonic mass of the monopole in kg. The numerical factor converts mass from kg to 
GeV in such a way that the units are consistent. We also define the dimensionless potential 
\be
\mathbb{V}=V {r_s^2\over m_p^2},
\ee
which, thanks to the definition  \eqref{change_H},  becomes
\be
\mathbb{V}\left(\chi\right)=\frac{\lambda_{\rm sm}}{4} m_p^2 r_s^2 \tilde{v}^4 \chi^2 \left(2+\chi\right)^2.
\ee
Finally, we define the dimensionless coupling function in an analogous way, as
\be
\mathbb{F}\left(\chi\right)=1+\xi \tilde{v}^2 \left(1+\chi\right)^2.
\ee

\subsection{Numerical integration method}

\noindent There exist different ways to perform the numerical integration of the eqs.\ \eqref{eom_r}-\eqref{KG_static}. We have chosen to treat these like an initial value problem (IVP), by integrating from the center of the body. We first find the internal solution and then use it at the boundary of the compact object to fix the initial conditions for the external solution. We choose to solve the system of  
equations with respect to the variables $\lambda$, $\nu$, $h$ and $p$ since $\rho=\rho_0$ is constant. In addition to the equations of motion, we must consider the TOV equation
\be
p_u=-\nu_u \left(p+\rho_0\right), 
\ee
where a subscript $u$ denotes a derivative with respect to $u$. 
In the top-hat profile approximation, this equation has the exact solution
\be
\frac{p}{\rho_0}=\mathcal{C} e^{-\nu} -1,
\label{TOV_int}
\ee
where $\mathcal{C}$ is a constant of integration that we can fix by the numerical shooting method. In order to 
find a guess for the shooting method, we use the standard GR expression for the pressure, see e.g. \cite{wald}. This is a very good approximation since we expect a small discrepancy between the GR solution and the numerical one, as explained in section \ref{sec4}. In our units, Eq.\ \eqref{TOV_int} becomes
\bea
{p(u)\over \rho_0}=\left[{\sqrt{1-s}-\sqrt{1-s^3 u^2}}\over \sqrt{1-s^3 u^2}-3\sqrt{1-s} \right].
\label{pressure_GR}
\eea
We impose the initial condition $\nu(u=0)=0$ \footnote{Note that $\nu(u=0)=0$  is not a regularity condition, which is instead given by asymptotic flatness, namely
 $\nu\left(u\rightarrow \infty\right)=0$. Since we solve an IVP, we prefer to fix $\nu(u=0)=0$ and then shift the solution to
$\nu(u)-\nu_{end}$, without loss of generality.}, so that we can write  $\mathcal{C}$ as 
\be
\mathcal{C}={p_c\over \rho_0}+1,
\ee
where $p_c=p(u=0)$ is the pressure at the center. We then optimize the value of $\mathcal{C}$ in such a way that it satisfies also the boundary condition for the pressure at the boundary $p(u=1/s)=0$. This method has also the advantage that it allows to test the limit for the central pressure  coming from GR \cite{wald}
\be
p(u=0)\longrightarrow\infty \Leftrightarrow \mathcal{R}={9m_b\over 4}.
\ee
It turns out that the difference between $\mathcal{C}$ in GR and in our model is so small that the two solutions are undistinguishable so we can effectively skip this step.
 
Therefore, we are left with the four equations  Eqs. \eqref{eom_r},  \eqref{eom_t},  \eqref{eom_theta}, and  \eqref{KG_static}. 
Of these, we keep  \eqref{eom_theta} as the Hamiltonian constraint and we integrate the other three as an IVP.
The initial conditions for $\lambda$, $\nu_u$ and $h_u$ are obtained from the regularity conditions of the solution at the center of 
the Higgs monopole
\bea
\lambda(0)=0,
\\
\nu_u(0)=0,
\\
h_u(0)=0.
\eea
In addition, we need to choose a value for  $h_c$ to begin the integration. We know that this value is an irrational number that can be determined numerically with finite accuracy only. Thus,  the basic idea of our algorithm consists of incrementing the value of $h_c$ digit by digit for a given number of digits. We also have an indication that makes integration easier. Indeed, we saw in section \ref{sec3} that if $h_{c}$ is larger than  $h^{\rm in}_{\rm eq}$ (see Eq.\ \eqref{h_eq_in}) then it never reaches the vev at spatial infinity. So, we can stop the integration as soon as $h>h_{\rm eq}^{\rm in}$ and reject the chosen value of $h_c$. Therefore, we begin by integrating from the 
approximate value of $h_{\rm eq}^{\rm in}$ (we recall that this value is calculated in the approximation that the internal solution is the same as GR) and if $h$ becomes larger than $h_{\rm eq}^{\rm in}$ we stop the integration, we keep the previous digit and perform once again numerical integration with a value of $h_c$ incremented by one less significant digit. Otherwise, namely when the 
Higgs field does not become higher than $h_{\rm eq}^{\rm in}$ and is trapped into the local minimum of the effective 
potential $h=0$, we increment the same digit. With this algorithm, we are able to maximize the precision on $h_c$ in 
the limit of the precision we impose or, in other words, we are able to push back the radial distance from the center of the body 
at which the scalar field is trapped into the local minimum of the effective potential $h=0$. 

We have also to take care of the ``degeneracy" of the  solution at spatial infinity. Indeed, 
the scalar field can tend to $\pm v$. So, when we perform numerical integration for different values of the parameters, we have to choose between the positive and  the negative solution.

\begin{figure}
\begin{center}
\includegraphics[scale=0.38, trim=280 0 310 0,clip=true]{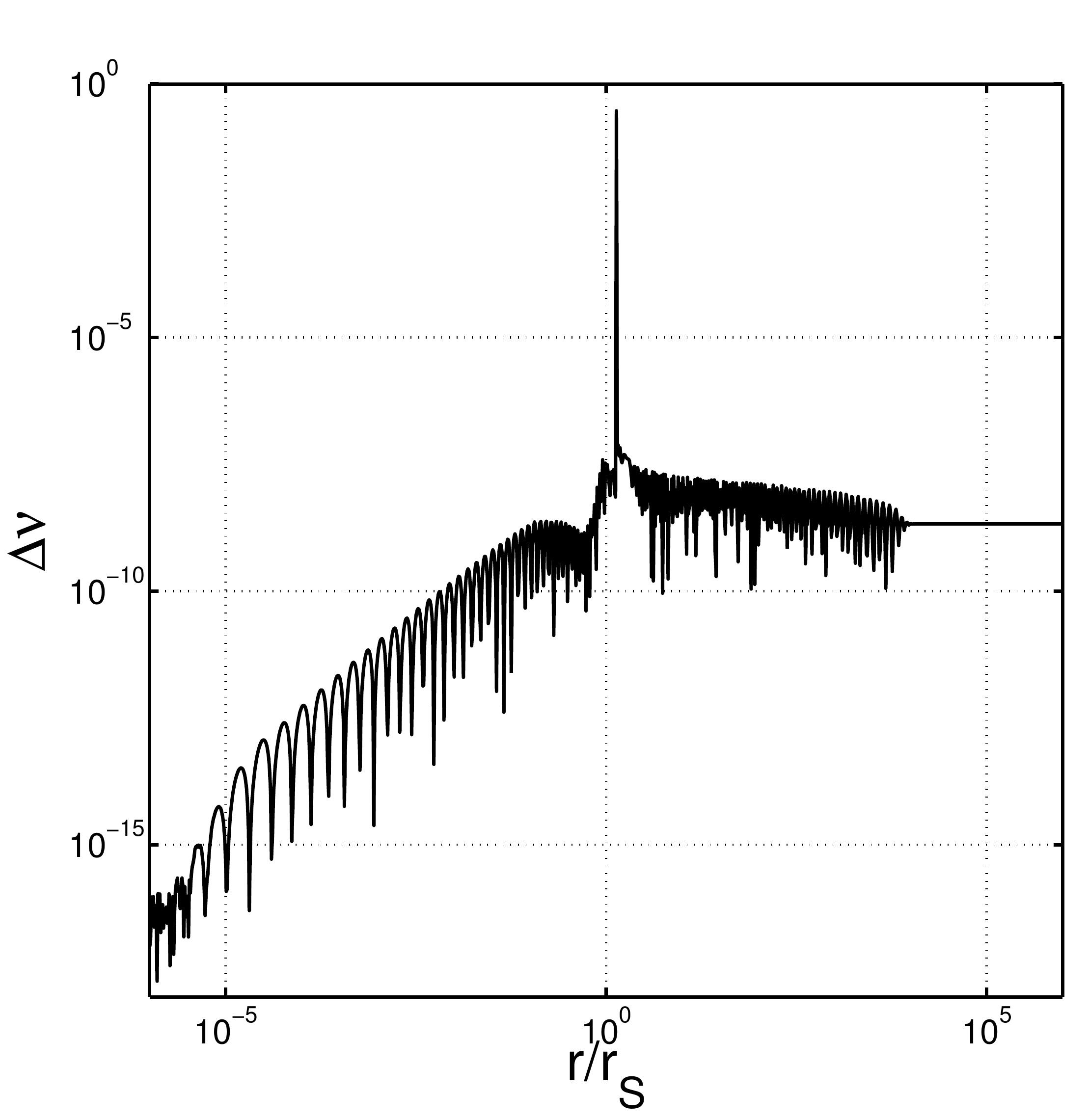}
\caption{
Hamiltonian constraint for the monopole solution ($\xi=10$, $m=10^6$kg and $s=0.75$) obtained with the full numerical method.
}
\centering
\label{plot_hamil_cons}
\end{center}
\end{figure}

In order to check the validity of our numerical code, we plot in Fig.\ \ref{plot_hamil_cons} the Hamiltonian constraint for the monopole solution presented in section \ref{sec5} and obtained with the full numerical integration. Here, the Hamiltonian constraint is defined as the absolute value of the difference between $\nu_u$ coming from Eq.\ \eqref{eom_theta},
where we replaced the values of all fields with the ones found numerically, and the value of $\nu_u$
determined numerically by solving the system of equations. The divergence appearing at the boundary of the monopole comes from the abrupt transition of energy density due to the top-hat approximation. Otherwise, the order of magnitude of the Hamiltonian constraint corresponds to what we expect from the numerical precision.

\subsection{Comparison between the full integration method and the simplified one}

\begin{figure}
\begin{center}
\includegraphics[scale=0.38, trim=280 0 310 0,clip=true]{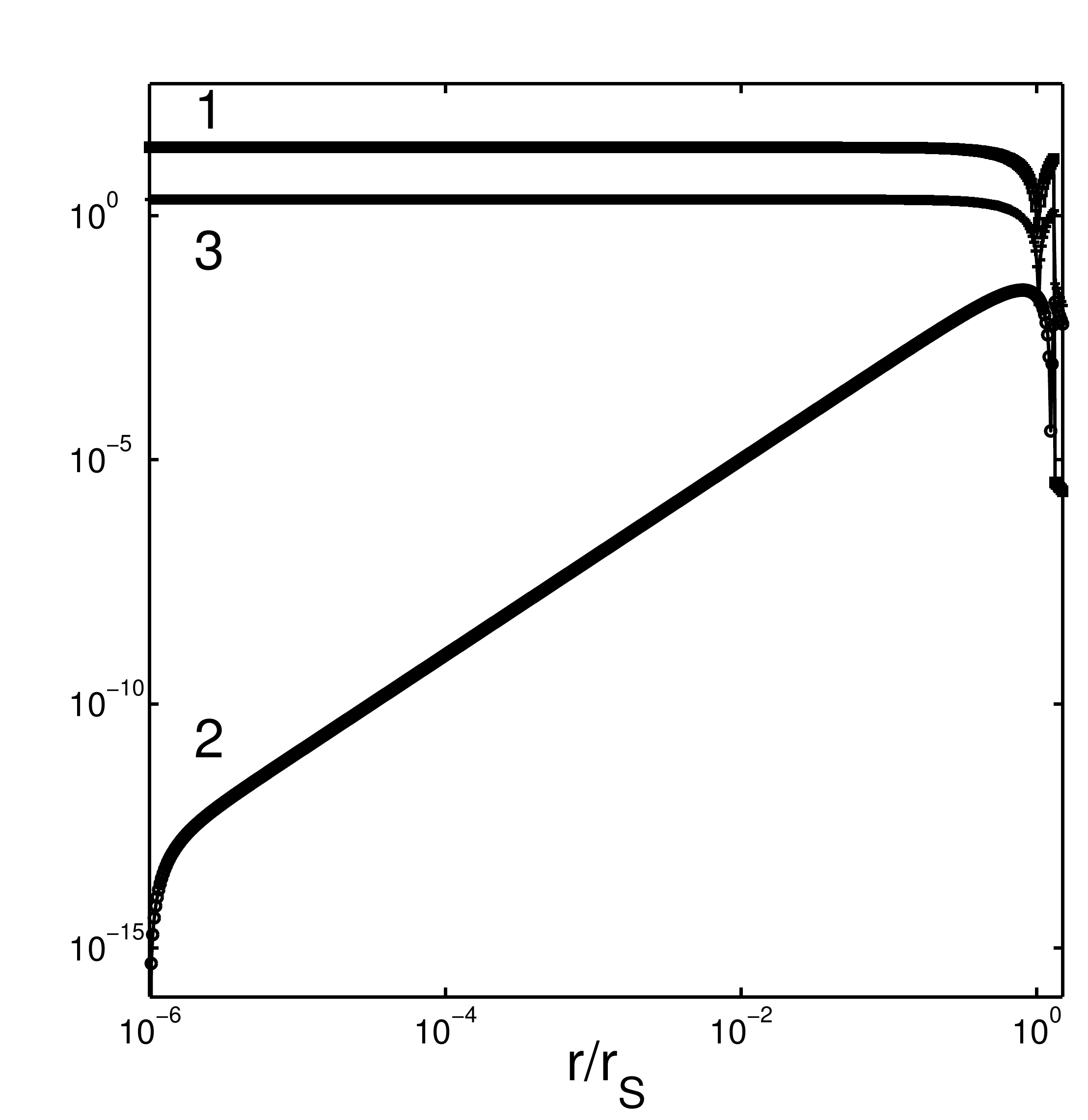}
\caption{Plot of  the trace of the energy-momentum tensor contributions $T^{(H)}$ (curve 2), $T^{(\xi)}$ (curve 3)
and of the left-hand side of Eq. \eqref{eom_tensor} (curve 1). The parameters are chosen as $m_b=10^6 \rm{kg}$, $s=0.75$, and $\xi=10$.}
\centering
\label{plot_trace_energy_momentum}
\end{center}
\end{figure}

\begin{figure}
\begin{center}
\includegraphics[scale=0.38, trim=280 0 310 0,clip=true]{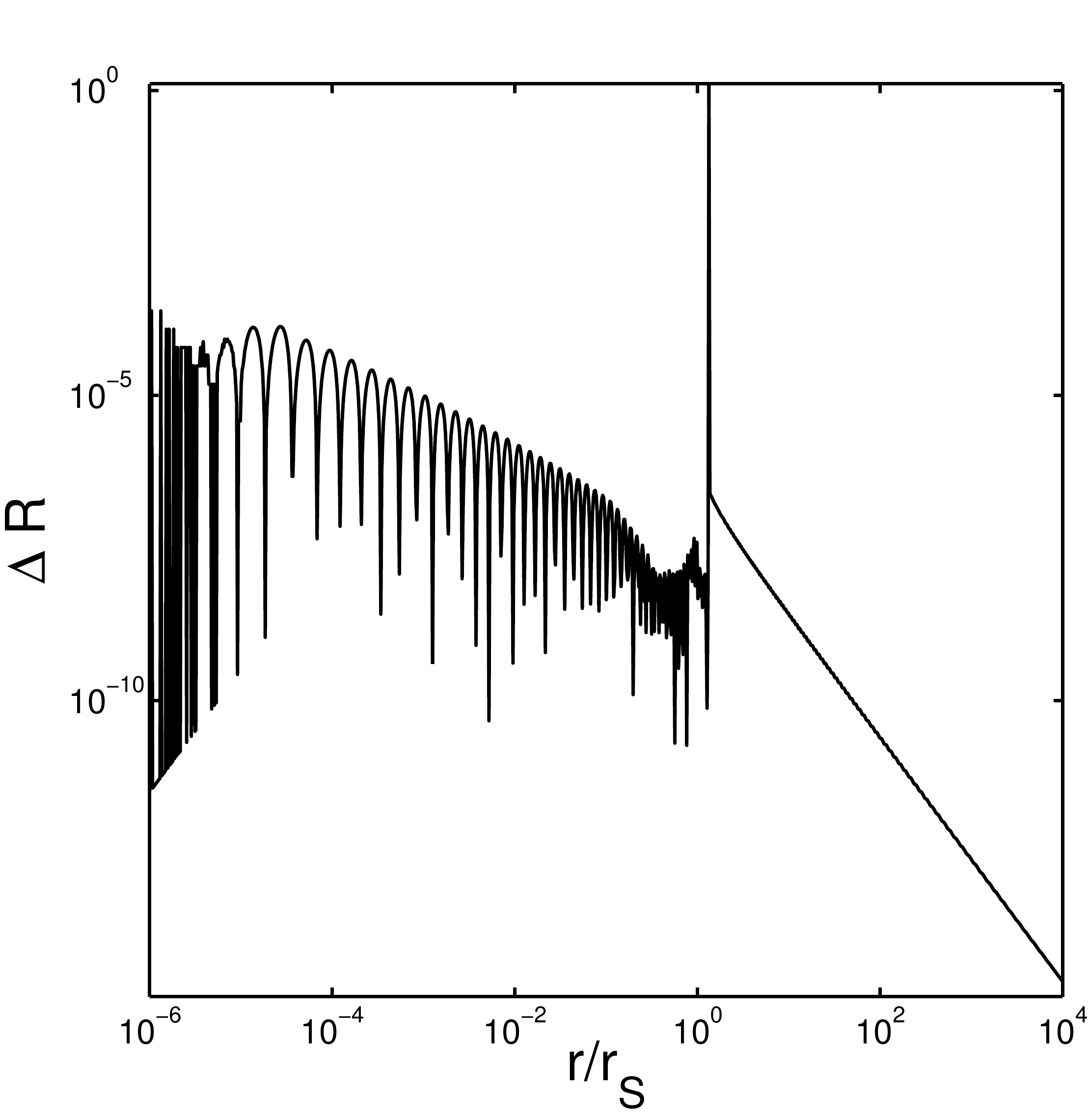}
\caption{Absolute value of the difference between the standard GR curvature scalar and its value calculated with our numerical algorithm for $m_b=10^6 \rm{kg}$, $s=0.75$, and $\xi=10$.
}
\centering
\label{plot_compare_scalar_curvature}
\end{center}
\end{figure}

\noindent In Sec.\ \ref{sec4} we argue that we can safely neglect the Higgs field inside the body as long as it is sufficiently constant 
and not too much displaced from its vev. This means that, instead of integrating the whole system of equations, we could use the inner Schwarzschild expressions for 
$\lambda$ and $\nu$ (Eqs.\ \eqref{nu} and \eqref{lam}) and integrate only the Klein-Gordon equation as an IVP. 
We now demonstrate the correctness of this claim by comparing our results with a complete numerical integration. We first plot on Fig.\ \ref{plot_trace_energy_momentum} the contribution of each term appearing in the trace of Eq.\ \eqref{eom_tensor} obtained with the full numerical integration in the case when $m_b=10^6 \rm{kg}$, $s=0.75$ and $\xi=10$. 
In the dimensionless unit system, the contributions of the trace of the energy-momentum tensor are given by
\bea
\frac{r_s^2}{m_{pl}^2} T^{(h)}=-h_u^2+2 V(h), 
\eea
and
\bea\non
\frac{r_s^2}{m_{pl}^2} T^{(\xi)}&\!\!=\!\!&\frac{3\xi}{4\pi} \Bigg\{h_u^2+h e^{-2\lambda}\Bigg[h_{uu}-h_u\left(\lambda_u-\nu_u-\frac{2}{u}\right)\!\!\Bigg]\!\Bigg\}.\\
\eea
We observe in Fig.\ \ref{plot_trace_energy_momentum} that the geometric part is clearly dominant while the contribution coming from the energy-momentum components of the scalar field is negligible. This result is confirmed by the comparison of the Ricci scalar given by Eq.\ \eqref{scalR} and the  expression  Eq.\ \eqref{analyticalR} evaluated numerically. In Fig.\ \ref{plot_compare_scalar_curvature} we plotted the absolute value of the difference between the two expressions  in function of the radial distance for the same parameters as in Fig.\ \ref{plot_trace_energy_momentum}. The difference is clearly negligible while the peak at the boundary of the body is caused only by the top-hat approximation for the energy density.

\begin{figure}
\begin{center}
\includegraphics[scale=0.38, trim=280 0 310 0,clip=true]{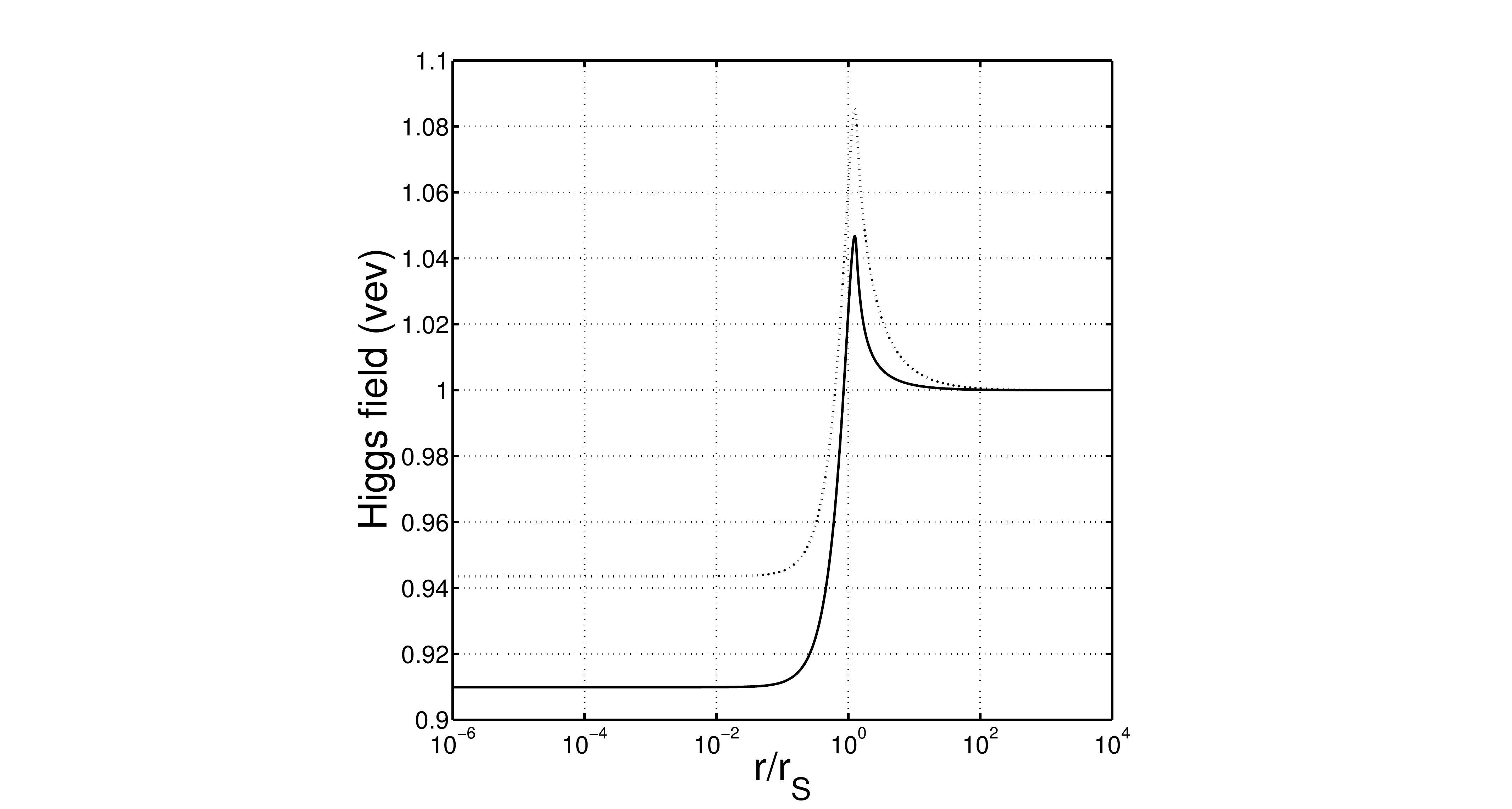}
\caption{
Numerical solutions for the monopole with $m_b=10^6 \rm{kg}$, $s=0.75$, and $\xi=10$ obtained with the full numerical integration and the simplified one.
The difference between the two solutions becomes apparent only inside the body and is negligible outside the body.
}
\centering
\label{plot_compare_num_int}
\end{center}
\end{figure}

\begin{figure}
\begin{center}
\includegraphics[scale=0.30, trim=340 0 0 0,clip=true]{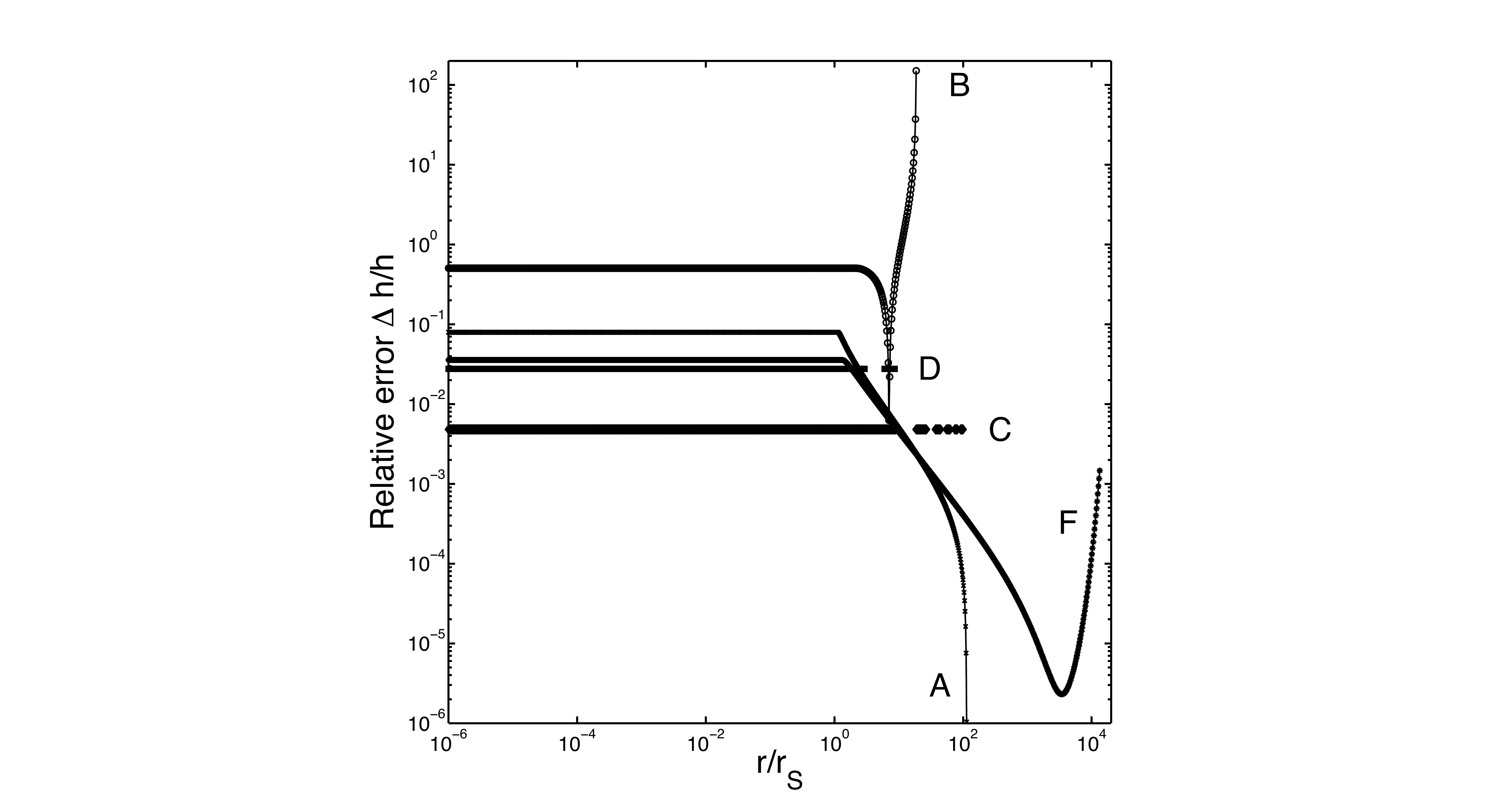}
\caption{
Relative error between the Higgs field solutions obtained with the full numerical method and the simplified one. 
Labels refer to Higgs monopoles of Table I.
}
\centering
\label{plot_rel_error}
\end{center}
\end{figure}

As a further check,  we plot the Higgs field profiles obtained with the two numerical methods in Fig.\ \ref{plot_compare_num_int}  for $\xi=10$, $m_b=10^6 \rm{kg}$, and  $s=0.75$. The discrepancy inside the body appears only because  the scalar field contribution is neglected in the simplified model. 
In order to get a quantitative result, we plot on Fig.\ \ref{plot_rel_error} the relative errors between the Higgs field solutions obtained with the full numerical method and the simplified one
for various monopole solutions. In general, we see that there is a very good agreement between numerical and approximate solutions only for small compactness.

\newpage


\end{document}